\documentclass[twocolumn]{aastex63}

\usepackage{txfonts}
\usepackage{graphicx}
\usepackage{aas_macros}

\begin{document}

\title{Impact of Dark Photon Emission on Massive Star Evolution and Pre-Supernova Neutrino Signal} 
\shortauthors{Sieverding et al.}
\shorttitle{Dark Photons in the late stages of Massive Star Evolution} 
\author[0000-0001-8235-5910]{A. Sieverding}
\correspondingauthor{A. Sieverding} 
\email{asieverd@umn.edu}
\affiliation{School of Physics and Astronomy,
      University of Minnesota, Minneapolis, MN 55455, USA} 
\author{E. Rrapaj}
   \affiliation{Department of Physics, University of California, Berkeley, CA 94720, USA} 
\affiliation{School of Physics and Astronomy,
      University of Minnesota, Minneapolis, MN 55455, USA} 
   
      \author[0000-0003-0859-3245]{G. Guo}
\affiliation{Institute of Physics, Academia Sinica, Taipei, 11529, Taiwan} 
\author[0000-0002-3146-2668]{Y.-Z. Qian}
\affiliation{School of Physics and Astronomy,
      University of Minnesota, Minneapolis, MN 55455, USA}

\begin{abstract}
     We study the effects of additional cooling due to the emission of a dark matter candidate particle, the dark photon, on the final phases of  the evolution of a $15\,M_\odot$ star and resulting modifications of the pre-supernova neutrino signal.
    For a substantial portion of the dark photon parameter space the extra cooling 
     speeds up Si burning, which results in a reduced number of 
     neutrinos emitted during the last day before core collapse. This reduction can be described by a systematic acceleration of the relevant timescales and the results can be estimated semi-analytically in good agreement with 
     the numerical simulations. Outside the semi-analytic regime
     we find more complicated effects. In a narrow parameter range, low-mass 
     dark photons lead to an increase of the number of emitted neutrinos because of additional shell burning episodes that delay core collapse. Furthermore, relatively strong couplings 
     produce a thermonuclear runaway during O burning, which could result in a complete
     disruption of the star but requires more detailed simulations to determine 
     the outcome. Our results
     show that pre-supernova neutrino signals are a potential probe of the dark photon parameter space.
\end{abstract}

\section{Introduction}
 
\label{sec:intro}
Dark matter, which represents more than 80\%~\citep{PDG2018} of the
matter density of the universe and whose nature remains one of the biggest mysteries in
physics, could be part of a dark sector which weakly interacts with Standard Model (SM) particles.
 Such scenarios of dark sectors naturally appear in many
extensions of the SM where dark matter particles only interact with the SM via a mediator. There is a rich experimental program searching for signatures of such mediators~\citep{Essig2013,Alexander:2016, Aaboud:2019}. However, if the coupling to SM particles is too weak, these particles could evade the searches and remain hidden. Astrophysical probes can greatly extend the reach of the search for dark matter candidates, trading the precision associated with the controlled environment of a laboratory for the vast range of densities and temperatures of stars~\citep{Raffelt:1996wa}. In one of the simplest extensions of the SM, the dark sector interacts with ordinary matter through the exchange of light vector bosons that couple to SM conserved currents~\citep{Holdom:1986,Rajpoot:1989,Nelson:1989fx,Batell:2014yra}. Dark matter is charged under a local U(1) symmetry in which the mediator couples to the SM electric charge
$Q$, and is described by the spin-one field $A^D_{\mu}$, called the
dark photon, which mixes kinetically with the standard
photon $A_{\mu}$.
Dark photons are characterized by two independent parameters, their mass $m_A$ and the reduced coupling strength with normal matter $\varepsilon$.
Several general dark matter searches have established experimental bounds on this parameter space \citep{Batell:2014,Essig2013}. 

Complementary constraints are provided by core collapse supernovae (SNe). For instance, parameters that would result in a noticeable reduction of the observed neutrino burst duration from SN 1987A can be excluded~\citep{Sung.Tu.ea:2019,Chang:2016,Rrapaj:2015,Raffelt:1990yz,Choplin.Coc.ea:2017}.
Additional constraints have been derived from observational signatures of our sun and other low-mass stars \citep{An.Pospelov.ea:2013}.  If beyond SM particles can be produced at relatively low temperatures, they would affect stellar evolution during He-burning and onward. This phase lasts millions of years and is sampled broadly by observations that can be used to derive stringent constraints \citep{Raffelt.Dearborn.ea:1987}.
Dark matter particles that are more massive, however, can only be produced at much higher temperatures and therefore only affect the short, advanced stages of the evolution of more massive stars. 

This work is a continuation of~\citet{Rrapaj.Sieverding.ea:2019} to study the potential impact of the dark photon on the final stages of a $15\,M_\odot$ star. During these stages temperatures in the star become high enough for dark photons to be emitted from electron-positron pair annihilation while the densities are still far below the regime where other processes, e.g., bremsstrahlung, become important. 
We therefore only include dark photons from pair annihilation in this work and
consider values of $m_A$ ranging from $2 m_\mathrm{e}$ to $10\,\mathrm{MeV}$, where $m_\mathrm{e}$ is the electron mass. Due to energy-momentum conservation, dark photons with masses below $2 m_\mathrm{e}$ cannot be produced by electron-positron pair annihilation and the temperatures needed for the emission of particles with masses more than $10\,\mathrm{MeV}$ are not reached before core collapse. We explore a wide range of coupling strengths $\varepsilon$, between $10^{-13}$ and $10^{-6}$, thereby probing a part of the parameter space unconstrained by neutrino observations of SN 1987A~\citep{Chang:2016,Hardy:2017}. 

If the dark photon is lighter than all other particles in the dark sector
it decays back into electron-positron pairs, either still inside the star or later in the interstellar medium.
In the latter case, the $\gamma$-ray signature of the subsequent annihilation of positrons can be used to derive constraints \citep{DeRocco2019}. If the dark photons decay inside the star, this may act as an additional heating mechanism leading to further constraints from the observed explosion energies of SNe \citep{Sung.Tu.ea:2019}. 
In contrast to these scenarios, we assume that the dark sector contains lighter particles that the dark photon ultimately decays into, as also discussed in \citet{Rrapaj.Sieverding.ea:2019}. In this case, any energy carried away by the dark photon is leaked into the dark sector, and the emission of the dark matter particles always acts as a cooling mechanism without additional signatures.
Under this assumption the constraints discussed by
\citet{DeRocco2019} and \citet{Sung.Tu.ea:2019} do not apply.

As a messenger of the stellar interior, we look at the emission of pre-SN neutrinos  and
 find that, in a large part of our selected dark photon parameter space, the extra cooling leads to a speed-up of the
 final burning stages and systematically reduces the number of neutrinos emitted during the last day before core collapse.
 We also show that this reduction can be estimated with good accuracy just based on the baseline stellar model by adjusting the time-integration of the neutrino luminosity. Our results suggest that this effect may be used 
to constrain the dark photon parameter space if pre-SN neutrinos are detected in the future.
In a very small region of the parameter space we also find that the effect of the extra cooling results in a slight increase of the neutrino emission.
In addition, for strong couplings unstable and explosive O burning occurs. While the latter case may potentially also provide constraints on the dark photon parameters, improved simulations are required to better determine the final outcome and observational signatures.

Our paper is organized as follows. In \S\ref{sec:calculations} we describe the setup for our calculations. In \S\ref{sec:standard} we discuss the details of our fiducial model of the evolution of a $15\,M_\odot$ star calculated without extra cooling. 
In \S\ref{sec:examples} we provide 
 an overview of the relevant parameter space and
discuss details of the three types of effects that we find. In \S\ref{sec:constraints} we briefly outline the possibility of deriving constraints from future observations of pre-SN neutrinos and summarize our results.

\section{Calculations}
\label{sec:calculations}
We implement a tabulation of the dark photon emission rates from \cite{Rrapaj.Sieverding.ea:2019} in the stellar evolution and hydrodynamics code KEPLER \citep{Weaver.Zimmerman.ea:1978, WOOSLEY:2007} and calculate the evolution of a $15\,M_\odot$ star with an initial composition of solar metallicity \citep{Lodders:2009}.
The mass loss prescription is based on \cite{Nieuwenhuijzen.DeJager:1990} and the mixing length for convection is equal to the pressure scale height. Semi-convection is treated as described in \cite{Woosley.Weaver.ea:1988}, limiting the convective diffusion coefficient to $10\%$ of the thermal value (see also \citealt{Heger.Woosley.ea:2002}). Overshoot and thermohaline mixing \citep{Kippenhahn.Ruschenplatt.ea:1980} are also included. Neutrino energy loss and the resulting total neutrino luminosity are based on \citet{Itoh:1996}.

We evolve the models until the onset of core collapse, which we define as the point when the infall velocity exceeds $5000\,\mathrm{km}/\mathrm{s}$.
This limiting value is higher than the value of $1000\,\mathrm{km}/\mathrm{s}$ used in previous studies \citep{WOOSLEY:2007,Heger.Woosley.ea:2002}
because the additional dark matter cooling tends to accelerate contraction.
\begin{figure}
    \centering
    \includegraphics[width=\linewidth]{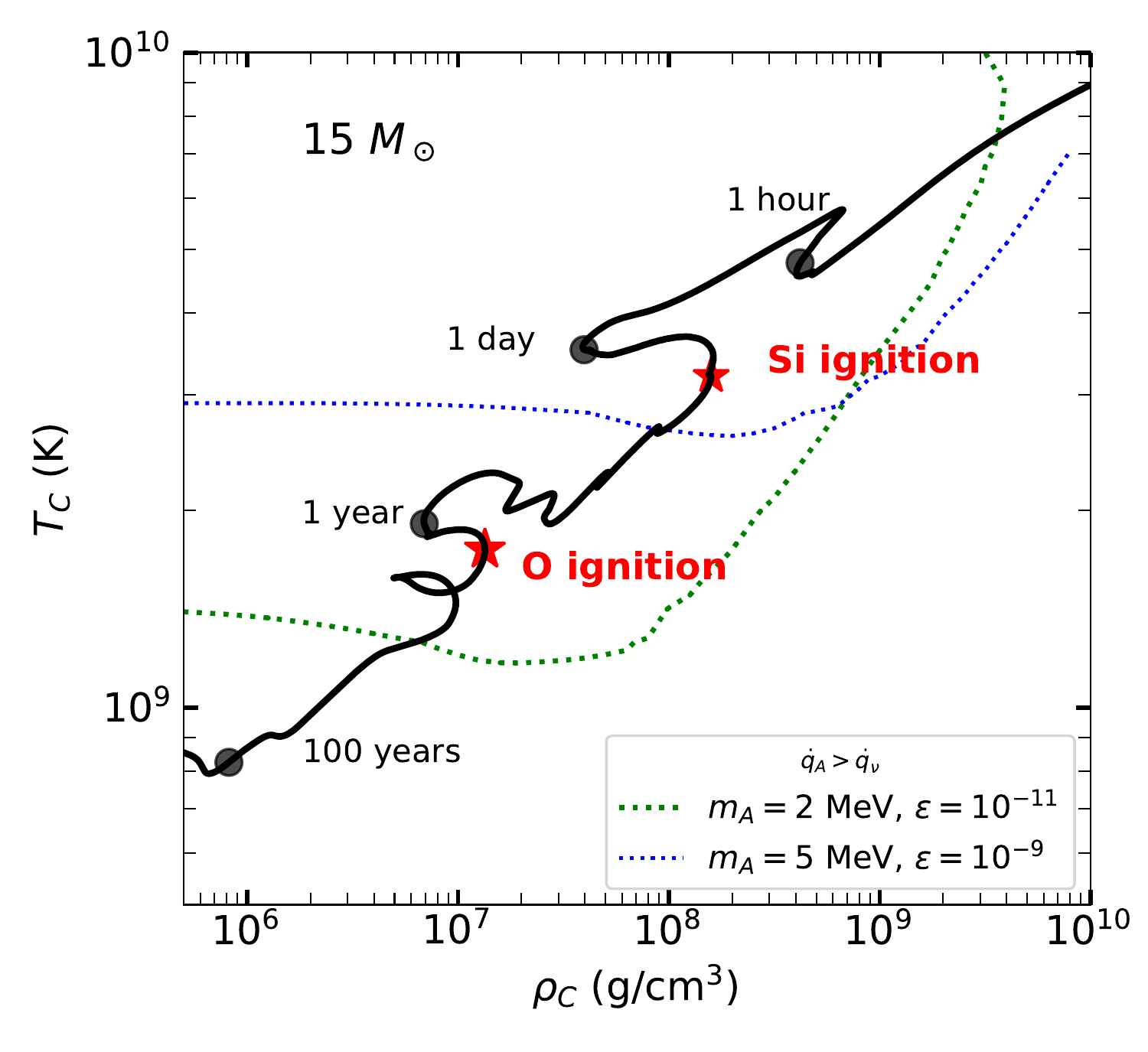}
    \caption{Track of core temperature vs. density for our fiducial stellar model without extra cooling. The time until core collapse is indicated at several points. In the region above the dotted contours for two sets of $m_A$ and $\varepsilon$, the corresponding dark photon emission dominates neutrino loss.}
    \label{fig:standard_track_s15}
\end{figure}
In order to explore the dark photon parameter space we have calculated more than 470 stellar models and to limit the computational cost
we only use an approximate 19-isotope nuclear reaction network to calculate the energy generation rates. At a temperature of 3.5 GK the code switches to solving  for the composition in quasi-statistical equilibrium (QSE) \citep{Hix.Khokhlov.ea:1998}. For even higher temperatures and when O is depleted full nuclear statistical equilibrium (NSE) is assumed. 
Furthermore, we limit the number of  zones in our models to 2000.

\section{Fiducial model}
\label{sec:standard}

Figure \ref{fig:standard_track_s15} shows the track of the fiducial stellar model without extra cooling in terms of central temperature $T_\mathrm{C}$ and density $\rho_\mathrm{C}$. This track is in good agreement with the results from \citet{WOOSLEY:2007} and also with those from different stellar evolution codes, such as MESA \citep{Paxton2015}. 
For the parameters that we are studying, dark photon production is not relevant for temperatures below $\sim 1$~GK. 
The dotted contours in Figure \ref{fig:standard_track_s15} indicate where energy loss due to dark photon emission equals the neutrino loss, marking the boundary of the temperature and density regime where dark photon loss would dominate. This boundary depends on the values of $m_A$ and $\varepsilon$. For instance, lighter dark photons with stronger coupling strengths start to dominate at lower temperatures. The details of these contours were discussed by \cite{Rrapaj.Sieverding.ea:2019}.
As illustrated for $m_A=2\,\mathrm{MeV}$ and $\varepsilon=10^{-11}$ in Figure \ref{fig:standard_track_s15}, dark photons may already become important right after central C depletion, which occurs around 64 years before core collapse. Therefore, for context, we provide a short description of the evolution after C depletion. 
\begin{figure}
    \centering
    \includegraphics[width=\linewidth]{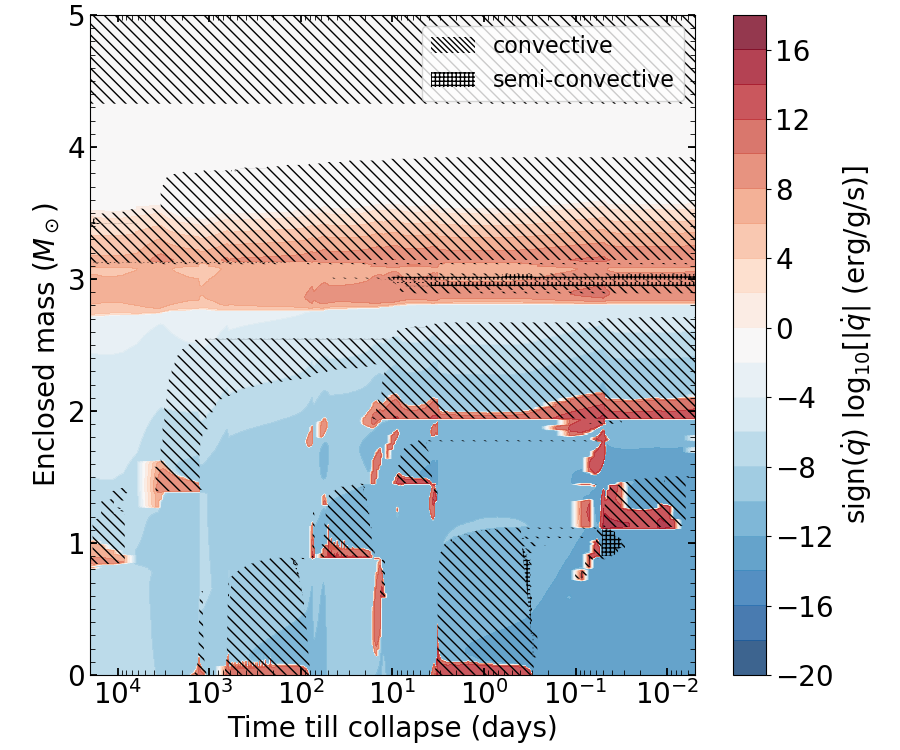}
    \caption{Kippenhahn diagram of the fiducial stellar model for the last 60 years before collapse. Hatched areas denote convective and semi-convective regions that are indicative of nuclear burning. Blue (red) background color indicates regions where cooling (nuclear energy generation) dominates.}
    \label{fig:kp_standard}
\end{figure}

The convective burning phases over the last 60 years are indicated in Figure \ref{fig:kp_standard}, where hatched areas indicate convection and the color code shows the net energy generation or loss. 
After C core and shell burning, Ne burning ignites at the center at $T_\mathrm{C}=1.35$~GK, around 1500 days before collapse under slightly degenerate conditions with a value of the degeneracy parameter $\eta=\mu_\mathrm{e}/k_\mathrm{B}T\approx 4$, where $\mu_\mathrm{e}$ is the chemical potential of electrons, $T$ the temperature and $k_\mathrm{B}$ the Boltzmann constant. The partial electron degeneracy leads to an initially rapid rise of the temperature, followed by core expansion at almost constant temperature. Once Ne is depleted, neutrino cooling leads to a decrease in temperature before the core continues to contract and eventually heats up again. This leads to the loop in the $T_\mathrm{C}$-$\rho_\mathrm{C}$ diagram shown in Figure \ref{fig:standard_track_s15}. Ne burning is visible as the small hatched peak in Figure  \ref{fig:kp_standard}, which also shows that this episode is very short.

About 650 days before collapse, O burning ignites centrally at $T_\mathrm{C}=1.7$~GK, which leads to a rapid rise of the central temperature and subsequent expansion. In contrast to the Ne-burning loop, at the end of O burning in the core the temperature does not decrease and the core is stabilized by shell burning. Once the shell burning ceases, neutrino loss reduces $T_\mathrm{C}$ while the core contracts. At higher densities, neutrino loss is suppressed and the track resumes its upward climb, while a second O-burning shell is ignited. Between core O depletion and Si ignition, He and C shell burning are still active.
\begin{figure}
    \centering
    \includegraphics[width=\linewidth]{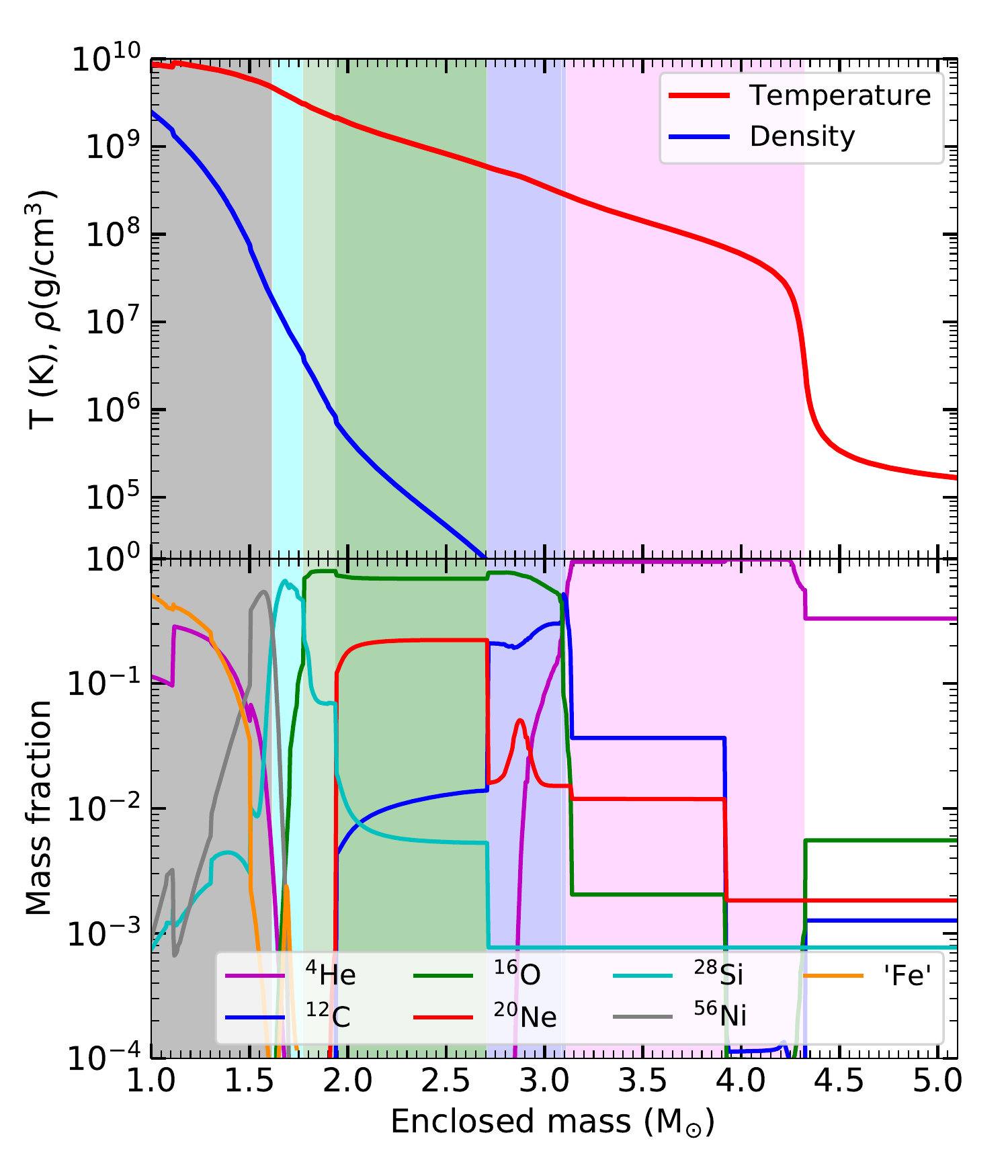}
    \caption{Stellar profiles of our fiducial 15 $M_\odot$ model at core collapse. The top panel shows temperature and density and the bottom panel mass fractions for the most important isotopes. }
    \label{fig:profile_standard}
\end{figure}
Convective Si burning ignites at the center around 3.5 days before collapse, after the second O-burning shell is extinguished.
Around 5.6 hours before collapse, central Si burning finishes and leaves a hot $1.1 M_\odot$ Fe core behind. This initial core grows further by shell burning that ignites about one hour before collapse and continues until the core mass exceeds its effective Chandrasekhar limit and collapses. 
Figure \ref{fig:profile_standard} shows the temperature, density, and mass fraction profiles of the most important isotopes of our fiducial model at core collapse.
The He core encompasses 4.26 $M_\odot$, the C/O core 3.02 $M_\odot$, the O/Ne core up to 2.70 $M_\odot$, and the Fe core encompasses 1.60 $M_\odot$ at a radius of $1336$ km. 

Multi-dimensional hydrodynamics simulations of the last minutes before core collapse \citep{Yadav.Naveen.ea:2020,Yoshida.Takiwaki.ea:2019,Mueller.Viallet.ea:2016} have shown deviations from spherically symmetric models using mixing length theory, but we do not expect our results to be qualitatively affected by these differences.

\begin{figure}
    \centering
    \includegraphics[width=\linewidth]{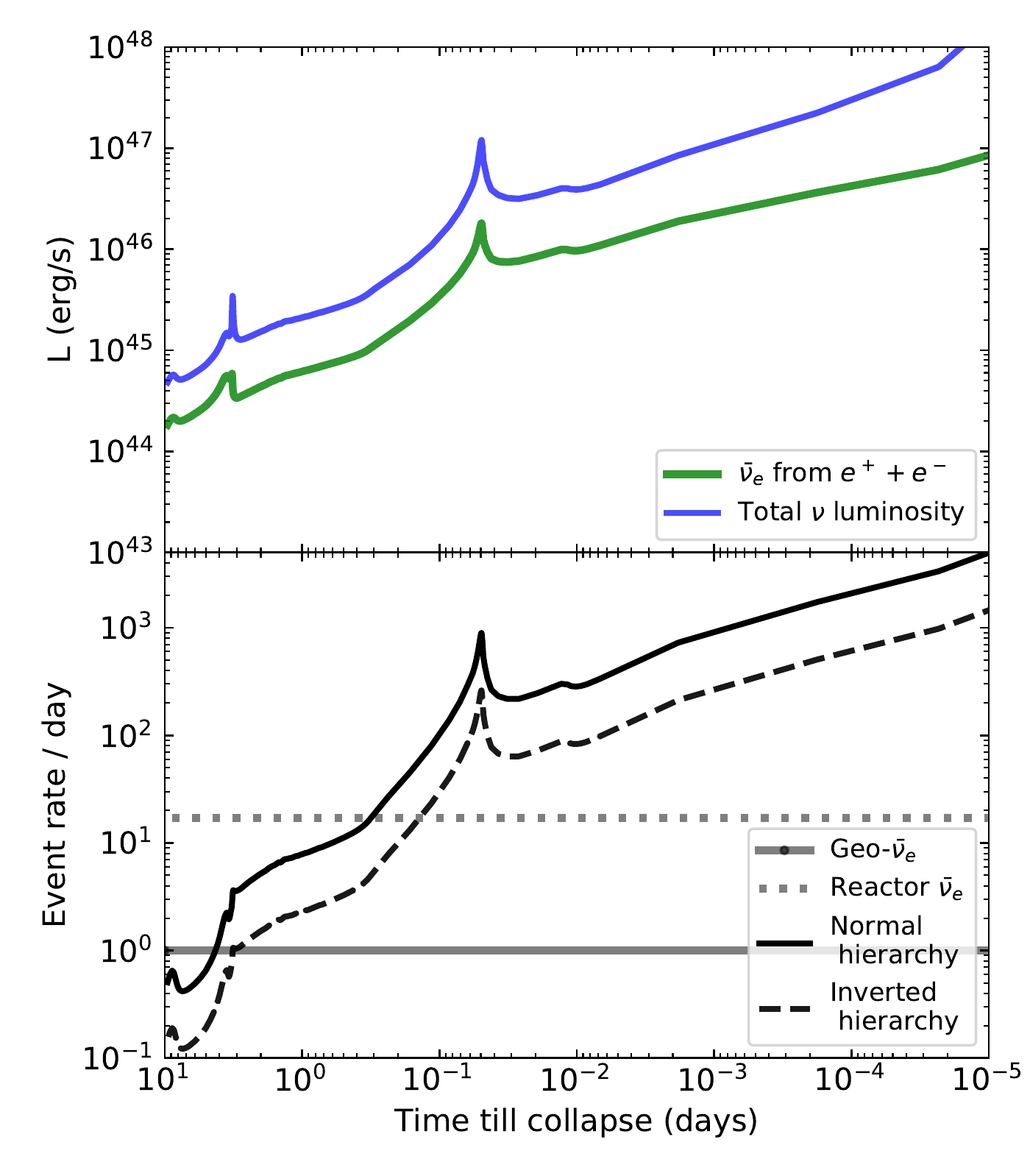}
    \caption{Top panel: Neutrino luminosities during the last 10 days before collapse from our fiducial model, without considering flavor transformations, for all neutrino species and processes as well as for only $\bar{\nu}_{\mathrm{e}}$ from pair annihilation. Bottom panel: Expected IBD event rates at JUNO for $\bar{\nu}_{\mathrm{e}}$ from pair annihilation assuming a distance of $500\,\mathrm{pc}$ to the source. The expected event rate depends on the neutrino mass ordering (normal or inverted hierarchy). The backgrounds from geo-neutrinos and nearby reactors are indicated by the horizontal gray lines.}
    \label{fig:count_rate_standard}
\end{figure}

Because the evolution of the stellar core is mostly decoupled from the surface during the final stages, neutrinos are unique messengers that may provide detailed information about the processes and conditions in the core shortly before collapse \citep[e.g.,][]{Guo.Qian:2016,Kato.Ishidoshiro.ea:2020}. 
Current and near-term neutrino detectors are expected to be able to detect the neutrinos from a nearby massive star only within the last day before collapse \citep[e.g.,][]{Guo.Qian.ea:2019,Kato.Ishidoshiro.ea:2020}. 
The top panel of Figure \ref{fig:count_rate_standard} shows the neutrino luminosity during the last 10 days before core collapse for our fiducial model.
In general, the luminosity increases as the star contracts and heats up, following the track in Figure \ref{fig:standard_track_s15}. 
The two peaks visible in Figure \ref{fig:count_rate_standard} are caused by the ignition of nuclear burning that leads to expansion and cooling, temporarily delaying collapse.
 The peak  at around 3.5 days before collapse corresponds to Si ignition at the center and
the peak at about one hour before collapse corresponds to the ignition of Si shell burning.

Here we focus on the $\bar{\nu}_{\mathrm{e}}$ for two reasons. Firstly, future scintillation detectors are likely to observe the pre-SN $\bar\nu_\mathrm{e}$, mainly by inverse beta decay (IBD, $\bar{\nu}_{\mathrm{e}}+\mathrm{p} \rightarrow \mathrm{e}^+ + \mathrm{n} $). Secondly,
towards the end of the life of a massive star 
 $\bar{\nu}_{\mathrm{e}}$ are mostly produced by electron-positron pair annihilation. In contrast to the emission of $\nu_\mathrm{e}$, this thermal process is relatively insensitive to the details of the stellar composition and does not depend on the uncertainties related to electron capture on nuclei. We calculate the spectral neutrino flux from pair annihilation as in \citet{Guo.Qian:2016},
using temperature, density, and electron fraction profiles as functions of time from our stellar models. 
The $\bar{\nu}_{\mathrm{e}}$ luminosity due to pair annihilation alone is shown in the top panel of Figure \ref{fig:count_rate_standard} in comparison to the total neutrino luminosity. At one day before collapse, $\bar{\nu}_{\mathrm{e}}$ constitute almost 30\% of the total luminosity. 
This fraction decreases towards collapse, as $\nu_\mathrm{e}$ from electron captures become increasingly important. At $10\,\mathrm{s}$ before collapse, however, $\bar{\nu}_{\mathrm{e}}$ from pair annihilation still account for $10\,\%$ of the total neutrino luminosity.

Based on the $\bar\nu_\mathrm{e}$ fluxes from the stellar evolution models, we calculate the expected pre-SN neutrino signal at the Jiangmen Underground Neutrino Observatory (JUNO) following \cite{Guo.Qian.ea:2019}.
With the spectral number luminosity $\phi_{\bar{\nu}_{\mathrm{e}}}(E_\nu,t)$ of the star, the expected energy-differential event rate at a distance $d$ is 
\begin{equation}
   \frac{d^2N}{dE_\nu dt} = \frac{1}{4 \pi d^2}\epsilon_{\mathrm{eff}} N_p \sigma_{\rm IBD}(E_\nu) \phi_{\bar{\nu}_{\mathrm{e}}}(E_\nu,t),
   \label{eq:rate}
\end{equation}
where $\epsilon_{\mathrm{eff}}=0.73$ is the detector efficiency and $N_p=1.45\times 10^{33}$ is the number of protons based on 20~kt of active detector material with a $12\,\%$ proton fraction \citep{An.Guangpeng.ea:2016}.
The IBD cross-section $\sigma_{\rm IBD}(E_\nu)$ is calculated as in  \cite{Guo.Qian:2016}. For the event rate we integrate Equation~(\ref{eq:rate}) over $E_\nu=1.8$--$4\,\mathrm{MeV}$, which is the optimal energy window for detection. 
Due to flavor transformations the detection rate depends on the neutrino mass ordering and the expected number of events for the normal hierarchy is about 3.4 times higher than for the inverted hierarchy \citep[e.g.,][]{Guo.Qian.ea:2019}. 
Figure \ref{fig:count_rate_standard} shows the expected event rates from our fiducial model assuming a distance of $500\,\mathrm{pc}$. We also show the main sources of background, geo-neutrinos and $\bar{\nu}_{\mathrm{e}}$ from nearby reactors. 
  For the inverted hierarchy about 15 events are expected during the last day, which are approximately the same as the background. 
 For the normal hierarchy, however, about 50 events are expected during the last day, which are a factor of $\sim 3$ more than the background.
Therefore, a pre-SN neutrino signal may be detected above the background during the final day before core collapse, especially in the case of the normal mass hierarchy.

\section{Results}
\label{sec:examples}
\begin{figure}
    \centering
    \includegraphics[width=\linewidth]{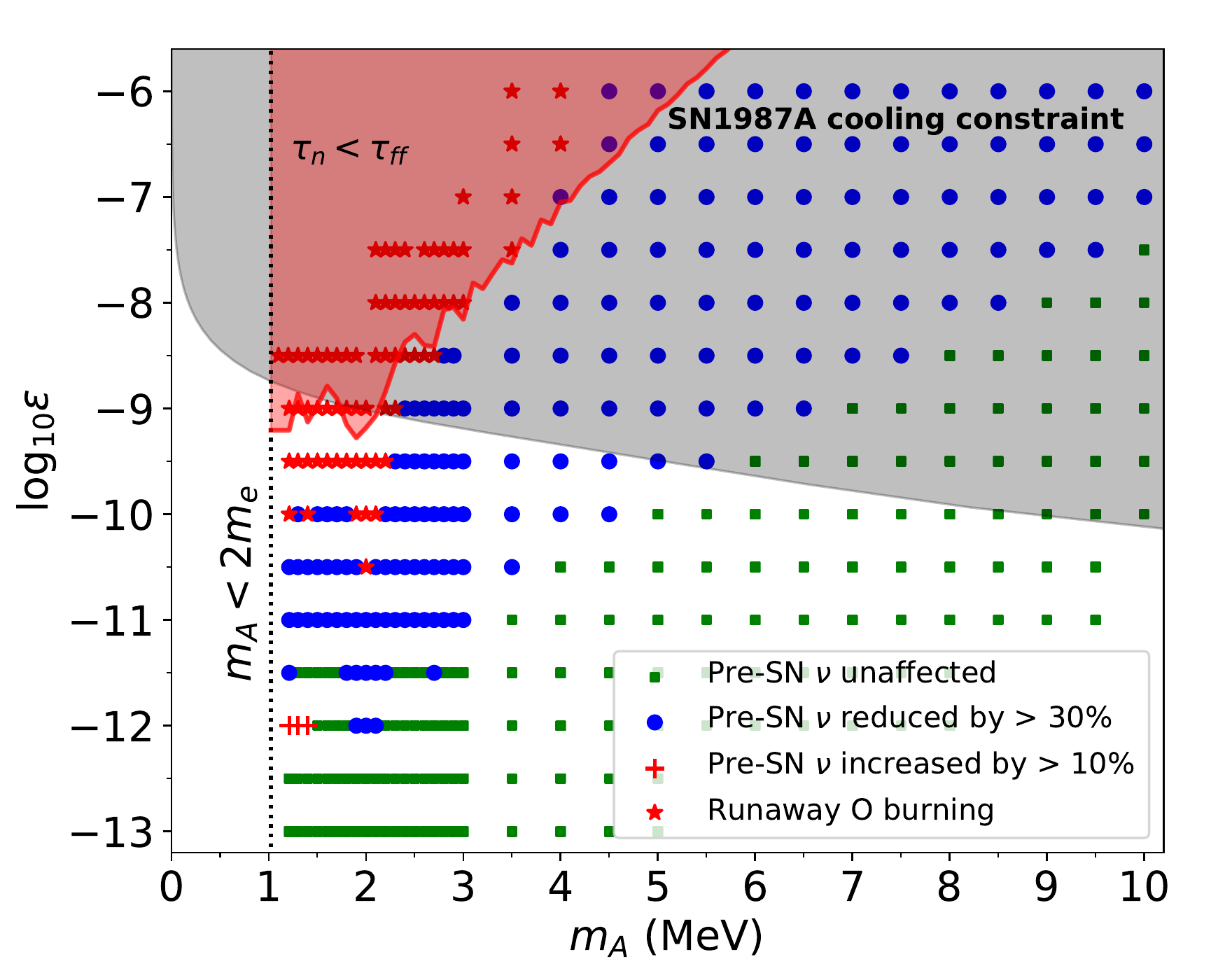}
    \caption{Overview of the results for a grid of the dark photon parameters $m_A$ and $\varepsilon$. The gray shaded area at the top indicates the parameter space already excluded by the detection of the neutrinos from SN 1987A \citep{Chang:2016}. Note also
    that we do not consider values of $m_A< 2\,m_\mathrm{e}$. The green squares mark models that are not noticeably affected by the dark photon emission. The blue circles mark models that exhibit a significant reduction of pre-SN neutrino emission (see \S\ref{sec:simple_reduction}), whereas the red crosses mark the three cases with noticeably increased neutrino emission (see \S\ref{sec:shell_effects}).
    For the red shaded area in the top left corner, runaway nuclear burning can be expected from simple arguments (see \S\ref{sec:O-deflagration}) 
    but full exploration is beyond the scope of this paper. The red stars indicate models that feature such explosive behavior.}
    \label{fig:overview}
\end{figure}

We have calculated stellar models for a grid of the dark photon parameters $m_A$ and $\varepsilon$ laid out in Figure \ref{fig:overview}.
For  $3\,\mathrm{MeV} < m_A < 10\,\mathrm{MeV}$ we have looked at  
increments of $0.5\,\mathrm{MeV}$ and for 
$2\,m_\mathrm{e}<m_A \leq 3\,\mathrm{MeV}$ we have taken smaller steps of  $0.1\,\mathrm{MeV}$. For $\varepsilon$ we looked at logarithmic intervals of $0.5\,\mathrm{dex}$ spanning values from $10^{-13}$ to $10^{-6}$.
For reference, the region of the parameter space that is already excluded by the detection of the neutrinos from SN 1987A \citep{Chang:2016} 
is indicated as the top gray shaded area in Figure \ref{fig:overview}. Note also that the region corresponding to $m_A<2\,m_\mathrm{e}$
is not studied here.
The different symbols in Figure \ref{fig:overview} indicate the outcomes of the models that we have calculated.

We find three qualitatively different ways in which the extra cooling changes the late phases of stellar evolution and the last-day pre-SN neutrino signal.
The green squares in Figure \ref{fig:overview} mark models that exhibit only negligible
deviations from the fiducial model. In contrast,
the blue circles mark cases in which the pre-SN neutrino emission is reduced by more than $30\,\%$ due to accelerated Si burning. 
Part of the parameter space with the blue circles is currently unconstrained, suggesting that future observations of pre-SN neutrinos may impose new constraints on the dark photon properties.

Only three models with small dark photon masses, indicated by the red crosses in Figure \ref{fig:overview}, exhibit a slight increase in the pre-SN neutrino emission. As we will show in \S\ref{sec:shell_effects}, this result is caused by 
an additional shell burning episode that delays core collapse. For sufficiently small dark photon masses and intermediate coupling strengths, the dark photon emission predominantly originates not from the stellar center, but from the region of the final the O/Ne shell. 

For the parameters in the red shaded region in the upper left corner of Figure \ref{fig:overview}, the nuclear timescale exceeds the hydrodynamic timescale during O burning and thermonuclear runaway is to be expected, as we will explain in \S\ref{sec:O-deflagration}. We did not perform full calculations to cover this whole region due to the complications associated with this explosive scenario. While the final fate of the corresponding models
requires further investigation beyond the scope of this paper, we can speculate that all these cases may even lead to thermonuclear SNe and the disruption of the whole star. 

In general, energy loss determines the timescales of the advanced burning stages in massive star evolution \citep{WOOSLEY:2007}.
The balance between energy generation due to nuclear burning and energy loss determines the temperature of hydrostatic burning and thus the rates at which nuclear reactions occur. Furthermore, energy loss determines how fast the stellar core can contract because part of the resulting gain in gravitational binding energy needs to be radiated away. 
Thus, emission of dark matter particles that increases the energy loss rate generally leads to a speed-up of stellar processes. The consequences of this acceleration, however, depend on where the speed-up occurs and on whether the star can adjust sufficiently fast.
Below we give more detailed explanations for the various effects that we observe.

\subsection{Reduced neutrino emission}
\label{sec:simple_reduction}
As long as the stellar path to core collapse is guided by energy loss, an additional cooling agent, such as dark photon emission, leads to a speed-up of the evolution. Once the core becomes hot enough, the bulk of the energy loss is taken over by the dark matter particles instead of the neutrinos, and the accelerated evolution leaves less time for neutrino emission.

\begin{figure}
    \centering
    \includegraphics[width=\linewidth]{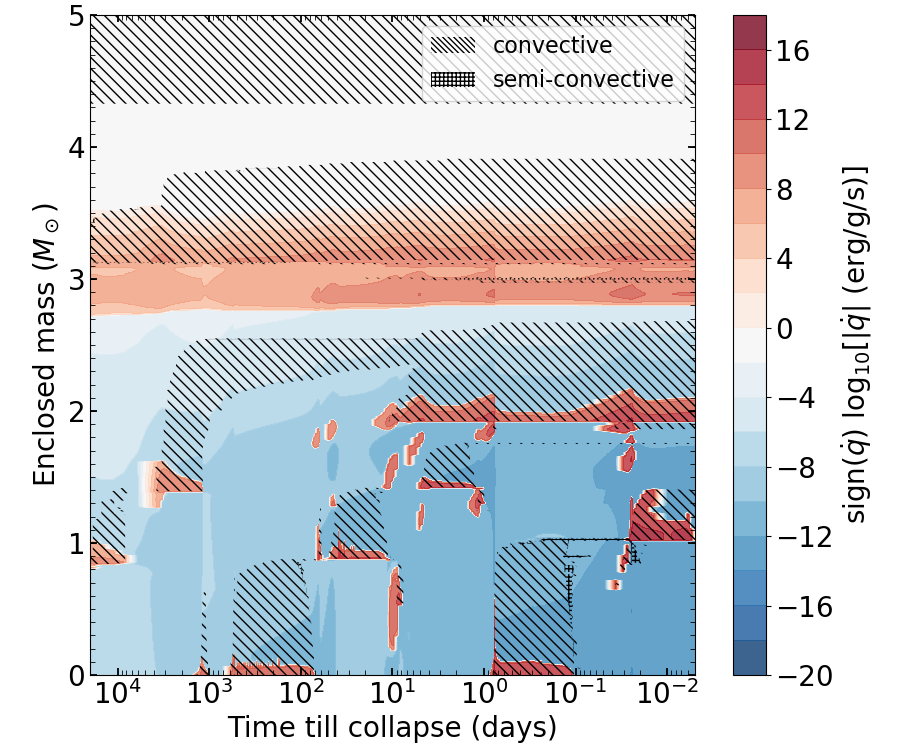}
    \caption{Kippenhahn diagram for the model with the parameters $m_A=2\,\mathrm{MeV}$ and $\varepsilon=10^{-12}$. The onset of convective Si burning in the core, indicated by the hatched region starting about 0.9 day before collapse, is much closer to core collapse than in the fiducial model without extra cooling shown in Figure \ref{fig:kp_standard}.}
    \label{fig:kp_2_24}
\end{figure}

As an example, Figure \ref{fig:kp_2_24} shows the convective burning phases for a calculation with the parameters $m_A=2\,\mathrm{MeV}$ and $\varepsilon=10^{-12}$. Comparison with Figure \ref{fig:kp_standard} shows that the stellar structure remains largely the same as in the fiducial model, but the evolution timescales change greatly.
The period between the onset of convective Si~burning and collapse is only around 0.9 day in Figure \ref{fig:kp_2_24}, compared to 3.5 days 
in the fiducial model (see Figure \ref{fig:kp_standard}). That between the end of convective Si burning at the center and collapse is
less than 3 hours, compared to 8 hours in the fiducial model.
The radial extent of the convective regions, on the other hand, remains almost identical. 
The slight reduction of the extent of the convective zone due to the more efficient energy loss, along with
the shorter time for shell burning due to the faster evolution of the core, results in a slightly smaller final Fe core mass of
$1.58\,M_\odot$, compared with $1.60\,M_\odot$ for the fiducial model.
With the remnant mass estimated by the point where the entropy decreases below $4\,k_\mathrm{B}/$baryon \citep{Heger.Woosley.ea:2002}, 
dark photon emission slightly reduces the potential neutron star mass from $1.78\,M_\odot$ to $1.76\,M_\odot$. 

\begin{figure}
    \centering
    \includegraphics[width=\linewidth]{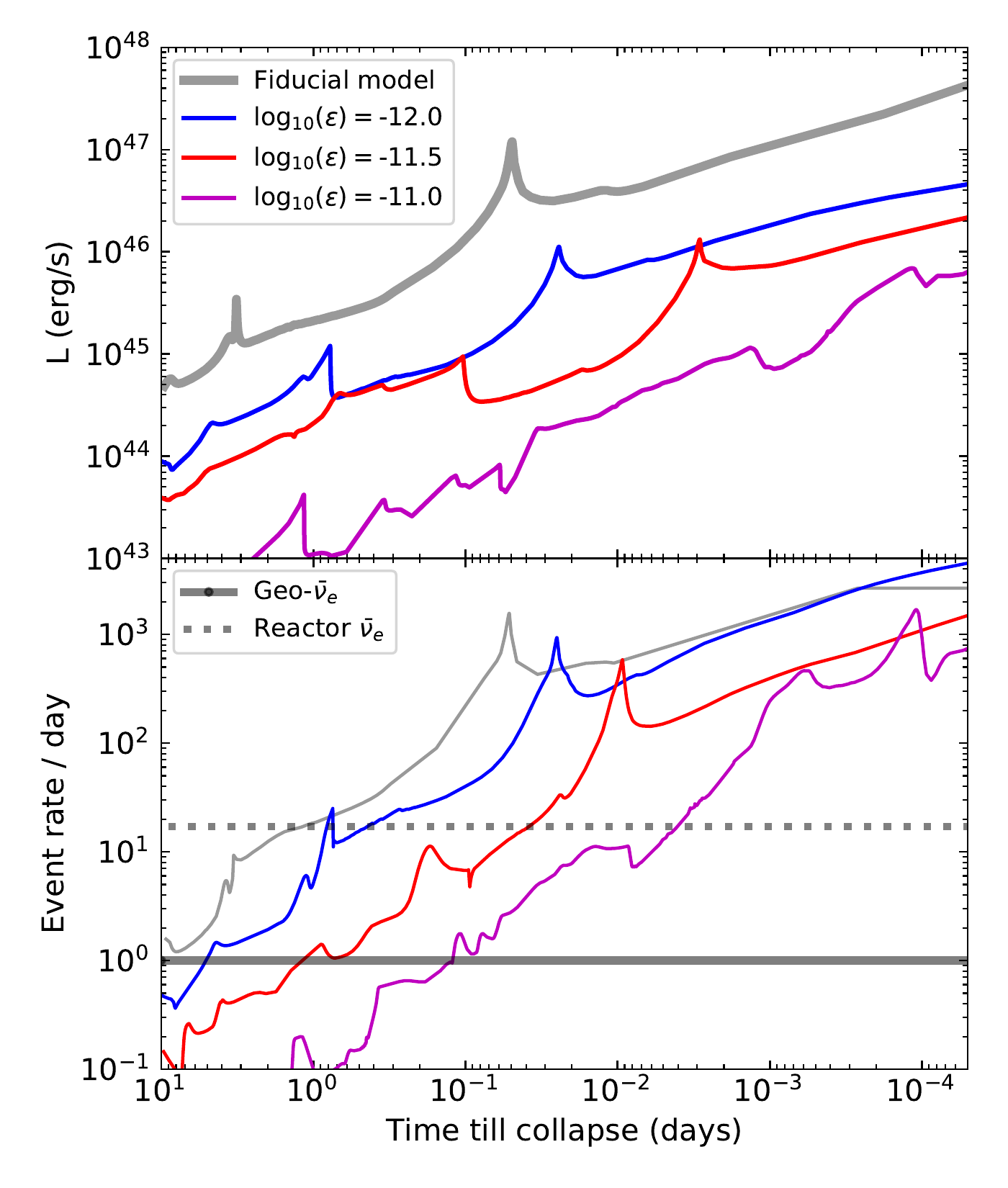}
    \caption{Evolution of $\bar{\nu}_{\mathrm{e}}$ luminosity from pair annihilation and associated event rate as in Figure \ref{fig:count_rate_standard}, but for models including dark photons with $m_A=2\,\mathrm{MeV}$. The event rate is calculated assuming the normal neutrino mass hierarchy and a distance of $500\,\mathrm{pc}$. With increasing values of $\varepsilon$ the luminosity is suppressed more and the peak associated with Si shell burning moves closer to the time of collapse, indicating the acceleration of the evolution.}
    \label{fig:lum_reduced}
\end{figure}

For the above values of $m_A$ and $\varepsilon$, the overall changes to the stellar structure relative to the fiducial model are minor. 
In contrast, the integrated number of neutrinos emitted during the last day is already decreased by a factor of two. 
This result shows that the main effect of the additional cooling is a speed-up of Si~burning and core contraction 
that scales with the energy loss rate. This speed-up of the evolutionary clock reduces the integrated neutrino 
emission with dark photons also contributing to the energy loss.
The effects on the $\bar{\nu}_{\mathrm{e}}$ luminosity from pair annihilation and the associated signal are illustrated in Figure \ref{fig:lum_reduced} 
for models with $m_A=2\,\mathrm{MeV}$ and a range of $\varepsilon$ values. In addition to a suppression of the luminosity, 
some features of the signal are shifted closer to the time of collapse. This shift is clearly visible for the peak 
associated with Si shell burning, which appears at about one hour before collapse for the fiducial model (see \S\ref{sec:standard}), 
but at only $30\,\mathrm{minutes}$ and $12\,\mathrm{minutes}$ before collapse for $\varepsilon=10^{-12}$ and $10^{-11.5}$, respectively. 
Note that for $\varepsilon=10^{-11}$, the evolution of the $\bar{\nu}_{\mathrm{e}}$ luminosity becomes less smooth due to complications discussed 
in \S\ref{sec:O-deflagration}. 

For all dark photon masses covered in our study, we find the same type of reduction of the neutrino emission, and the necessary coupling strength to achieve the same reduction increases with mass. 
In the following we show that this reduction can be estimated semi-analytically, just based on the fiducial stellar model. 
Without dark photon cooling, the timescales during the final stages of evolution are determined mostly by the neutrino loss. 
We expect $dt\propto L_\nu(t)^{-1}$, where $L_\nu(t)$ is the total neutrino luminosity from the whole star at time $t$. With the additional dark photon
cooling, we expect
\begin{equation}
dt'=\frac{L_\nu(t)}{L_\nu(t)+L_{A}(t)}dt,
\end{equation} 
where $L_A(t)$ is the dark photon luminosity from the whole star.
This change affects the results of any time-integrated quantity in two ways. 
Firstly, the time measure itself is changed. Secondly, the effective integration limits are changed.
The time of core collapse, $t_{\mathrm{CC}}$, marks the final time with a well-defined physical condition. In order to accumulate a fixed time interval $\Delta t$, e.g., one day, up to $t_{\mathrm{CC}}$, the starting point needs to be modified such that
\begin{equation} 
\Delta t=\int \limits_{t_0}^{t_{\mathrm{CC}}} dt =\int\limits_{t_0'}^{t_{\mathrm{CC}}} dt'.
\end{equation}
With $dt'<dt$ due to dark photon cooling, we have $t_0'<t_0$, so $\Delta t$ samples earlier stages of the evolution. 

The above prescription allows us to estimate the energy emitted in $\bar\nu_\mathrm{e}$ from pair annihilation during the last day before collapse as
\begin{equation} 
\label{eq:semi_analytic}
E_{\mathrm{1 day}}'=\int\limits_{t_0'}^{t_{\mathrm{CC}}} L_{\bar{\nu}_{\mathrm{e}},\mathrm{pair}}(t) \left( \frac{L_\nu(t)}{L_\nu(t)+L_{A}(t)} \right)  dt,
\end{equation}
where $L_{\bar{\nu}_{\mathrm{e}},\mathrm{pair}}(t)$ is the $\bar\nu_\mathrm{e}$ luminosity due to pair annihilation from the whole star. Note that $L_\nu(t)$ and $L_{A}(t)$ are based on the temperature and density profiles of the fiducial model without dark photon emission. This approximation is valid as long as the effects 
of extra cooling on stellar structure are negligible. In this case, the change of the $\bar\nu_\mathrm{e}$ emission can be estimated semi-analytically without computing new stellar models with the additional energy loss. 

\begin{figure}
    \centering
    \includegraphics[width=\linewidth]{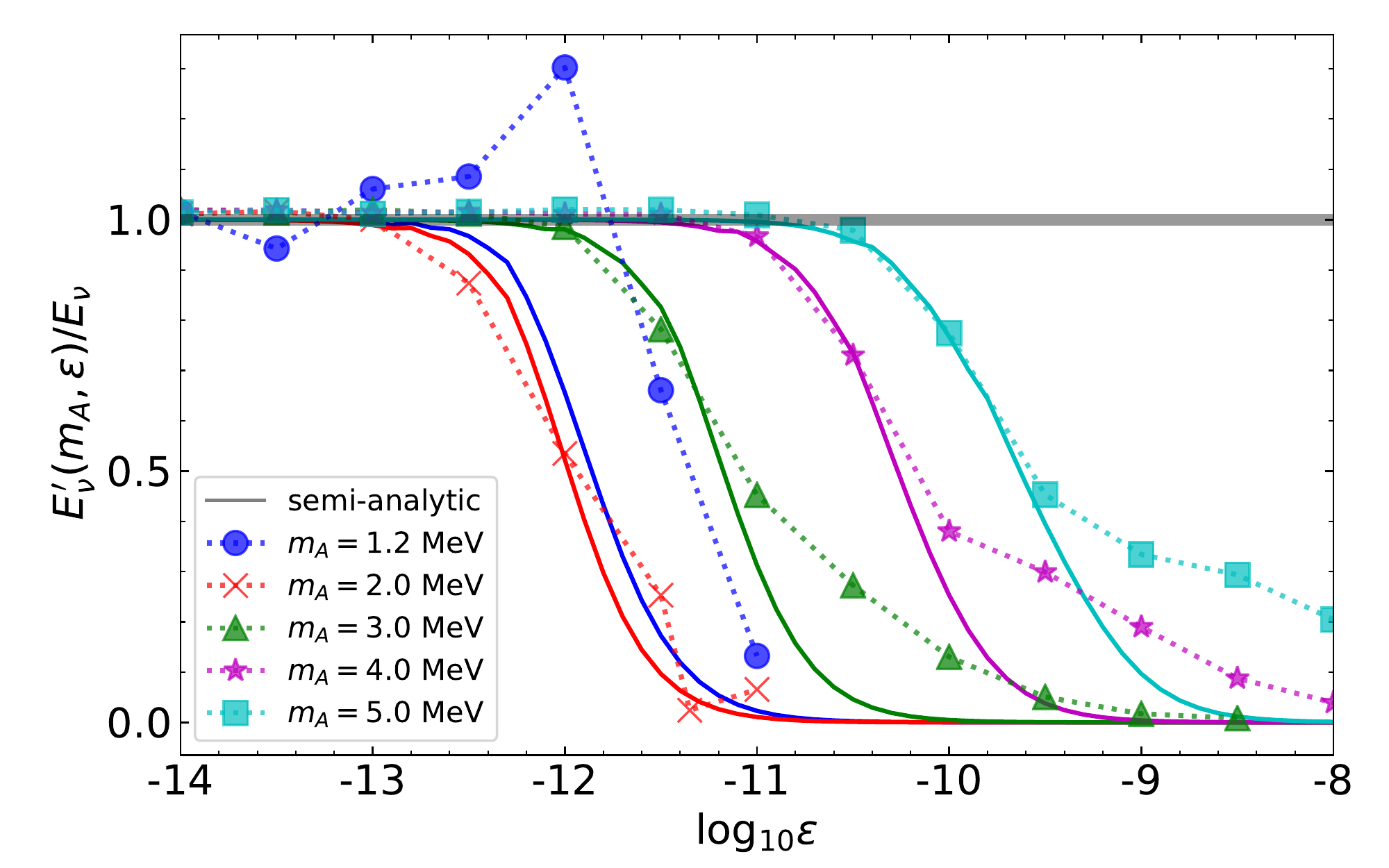}
    \caption{Energy 
    carried away by $\bar{\nu}_{\mathrm{e}}$ over the last day before collapse relative to the fiducial model.
    Semi-analytic estimates based on Eq. (\ref{eq:semi_analytic}) are displayed as solid curves and compared to the results from full simulations (symbols connected with dotted line segments). The gray horizontal bar indicates no change from the fiducial model.}
    \label{fig:prediction_vs_models}
\end{figure}

Figure \ref{fig:prediction_vs_models} displays the energy emitted in $\bar{\nu}_{\mathrm{e}}$ from pair annihilation over the last day before core collapse, normalized to the fiducial model, as a function of the dark photon coupling strength for a range of $m_A$.
The symbols show the results from full stellar evolution calculations and demonstrate that increasing dark photon emission leads to a reduction of the neutrino emission. The solid curves show the estimates based on Eq.~(\ref{eq:semi_analytic}),
which successfully describe the onset of the effects of dark photons as well as the overall trend for many of the calculated models,
especially when the reduction of the neutrino emission is $\lesssim50\%$. Deviations from the actual calculations are expected
when the dark photon cooling starts to change the temperature and density profiles for sufficiently large values of $\varepsilon$.
For a large part of the dark photon parameter space, however, the main effect is the relatively straightforward reduction of the 
neutrino emission without significant changes to the stellar structure as discussed above. Those cases are indicated by 
the blue circles in Figure \ref{fig:overview} and cover an unconstrained region of the parameter space.

Figure \ref{fig:prediction_vs_models} also shows large deviations of the estimates from the actual calculations 
for $m_A=1.2$~MeV. As we will explain in \S\ref{sec:shell_effects}, these deviations are
due to changes of the shell-burning evolution.

\subsection{Increased neutrino emission}
\label{sec:shell_effects}
\begin{figure}
    \centering
    \includegraphics[width=\linewidth]{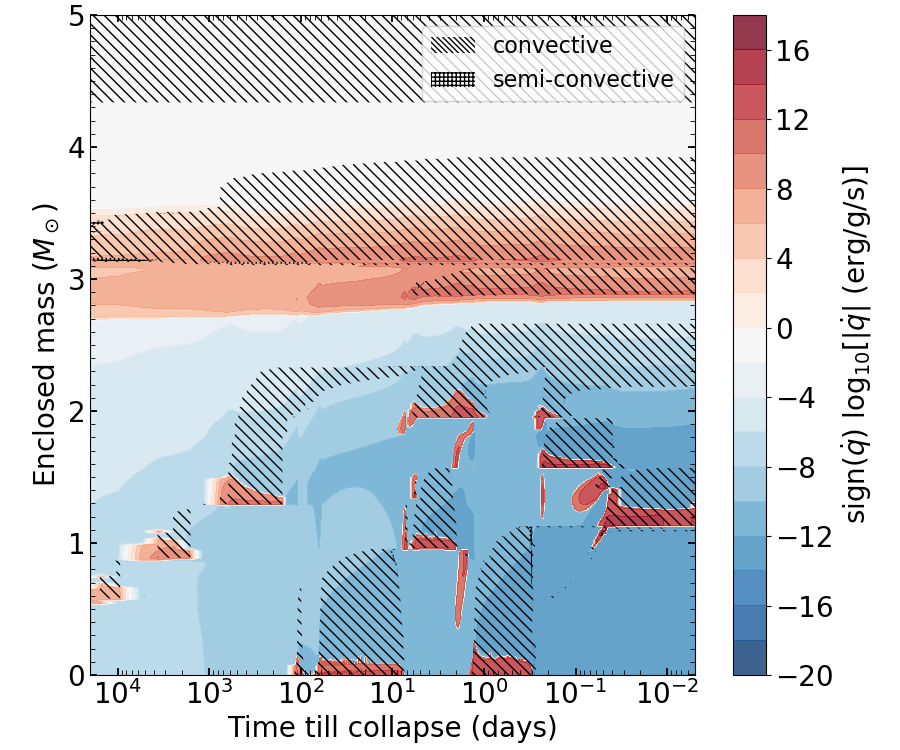}
    \caption{Kippenhahn diagram for the model with the parameters $m_A=1.2\,\mathrm{MeV}$ and $\varepsilon=10^{-12}$. At 0.25 days before collapse, an additional convective O burning shell ignites between the enclosed mass coordinates of $1.6\,M_\odot$ and $2\,M_\odot$, which delays core collapse and allows more time for neutrino emission.}
    \label{fig:kp_121_24}
\end{figure}

As discussed in \S\ref{sec:simple_reduction}, the reduction of the pre-SN neutrino emission results almost entirely from the modification of the timescales of core burning and contraction. Below we show that the deviations from this explanation for low $m_A$ (see Figure \ref{fig:prediction_vs_models})
are due to the appearance of shell burning phases with a stabilizing effect, which leads to the cases of the increased neutrino emission
marked by the red crosses in Figure \ref{fig:overview}. 

For the dark photon parameters under our consideration, only three models with $m_A=2\,m_\mathrm{e}$, $1.1\,\mathrm{MeV}$, and $1.2\,\mathrm{MeV}$, all with $\varepsilon=10^{-12}$, have an increase of $\lesssim 50\%$ in the pre-SN neutrino emission relative to the fiducial model.
The comparison of the neutrino emission with the fiducial model is shown for the case of $m_A=1.2\,\mathrm{MeV}$ in Figure \ref{fig:prediction_vs_models}.
The Kippenhahn diagram for this case is shown in Figure \ref{fig:kp_121_24}, which reveals 
an additional episode of convective O shell burning between an enclosed mass of $1.6\,M_\odot$ and $2\,M_\odot$.
This episode starts at $0.25\, \mathrm{days}$ before collapse, following the end of core Si burning, but before the onset of Si shell burning.
Such an episode occurs neither in the fiducial model nor in the model with the reduced neutrino emission shown in Figure \ref{fig:kp_2_24}.
Heating by this additional O-shell burning phase relieves the pressure on the Si-core and delays the ignition of Si-shell burning and hence core collapse. This delay gives the star more time, thereby increasing the total number of pre-SN neutrinos emitted.

For $m_A=1.2\,\mathrm{MeV}$, the additional O-shell burning phase also occurs for larger values of the dark photon coupling,
but it cannot compensate for the effects of the accelerated burning and core contraction. Consequently,
the integrated neutrino emission for such cases is still reduced (see Figure \ref{fig:prediction_vs_models}), 
although not by as much as estimated from Eq. (\ref{eq:semi_analytic}),
which does not take into account the additional O-shell burning. 
We can only find a net increase of the pre-SN neutrino emission for a very narrow range of $\varepsilon$, which is strong enough to cause an additional burning episode, but weak enough to limit the acceleration of the evolution.

\begin{figure}
    \centering
    \includegraphics[width=\linewidth]{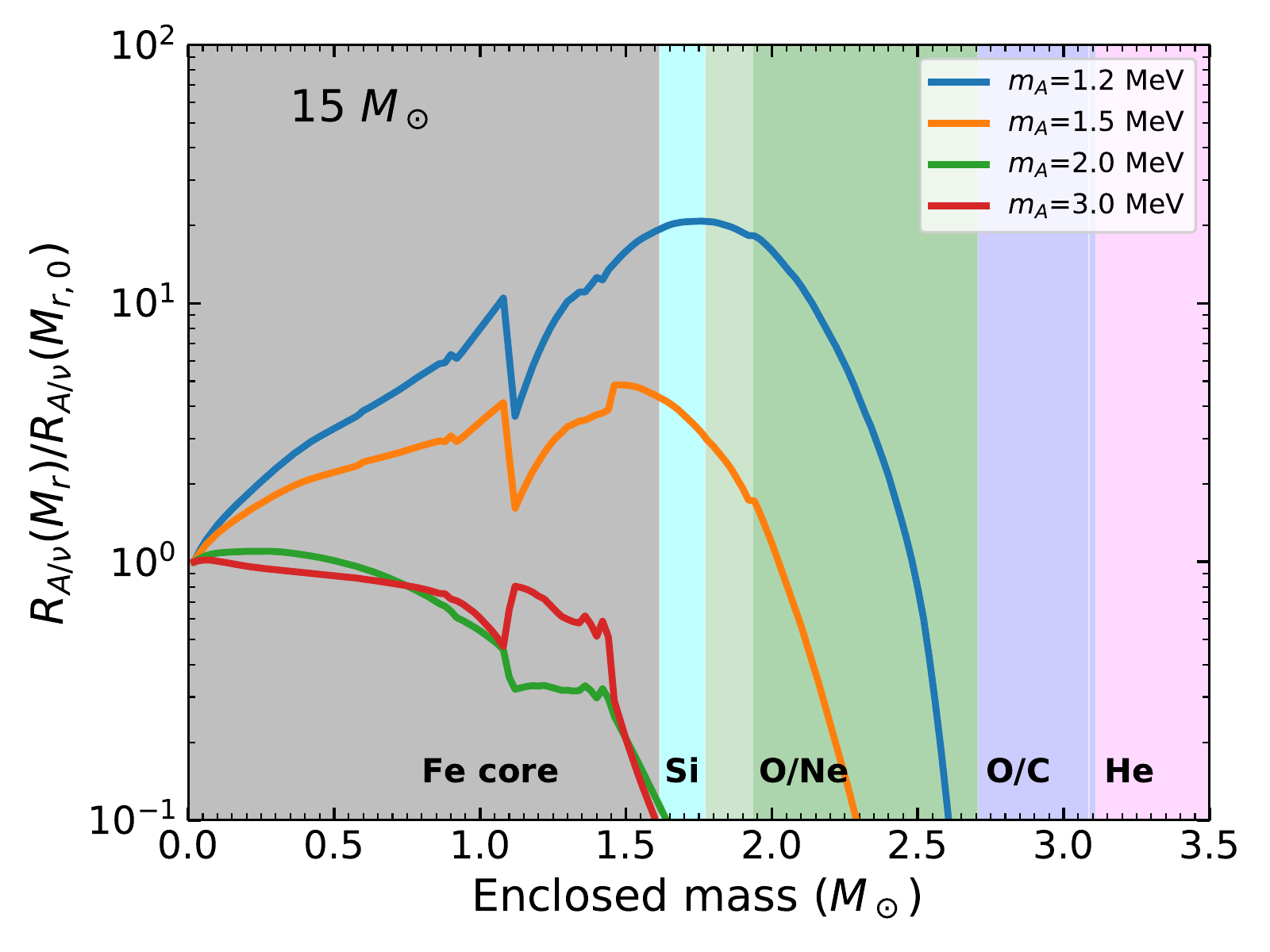}
    \caption{Ratio of dark photon to neutrino energy loss rates $R_{A/\nu}(M_r)$ (normalized to the central zone) as a function of the enclosed mass for the fiducial stellar model. The background colors indicate the shells with different compositions shown in Figure \ref{fig:profile_standard}.
    For $m_A<2\,\mathrm{MeV}$, the strongest dark photon emission originates not from the Fe core, but from the Si and O/Ne shells due to the suppression at high densities.}
    \label{fig:ratio_profile}
\end{figure}

In order to understand the apparent association of the additional burning episodes with the lightest dark photons in our study, we look at the time-integrated energy loss as a function of the enclosed mass $M_r$ inside radius $r$,
\begin{equation}
    E_{\mathrm{loss},x}(M_r)=\int\limits_{t_0}^{t_{CC}} \dot{q}_x(M_r,t) dt,
\end{equation}
where $x$ is either $\nu$ for the neutrino with the energy loss rate $\dot{q}_\nu(M_r,t)$ or $A$ for the dark photon with the energy loss rate $\dot{q}_A(M_r,t)$, and $t_0$ is the time of C ignition when the energy loss due to neutrinos and dark photons becomes relevant. 
Figure \ref{fig:ratio_profile} shows the ratio $R_{A/\nu}(M_r)=E_{\mathrm{loss,A}}(M_r)/E_{\mathrm{loss},\nu}(M_r)$ relative to the value for the central zone $R_{A/\nu}(M_{r,0})$ based on the fiducial model. Due to the scaling of $R_{A/\nu}(M_r)$ with $\varepsilon$, the quantity $R_{A/\nu}(M_r)/R_{A/\nu}(M_{r,0})$ is independent of $\varepsilon$. It is, however, quite sensitive to $m_A$.

For masses above $2\,\mathrm{MeV}$ the emission of dark photons dominates that of neutrinos mostly in the final Fe core because the temperature is generally higher at smaller radii. 
For masses below $2\,\mathrm{MeV}$, however, the dark photon emission dominates further outside the core. 
This different behavior for the lower values of $m_A$ can be understood as follows.
Because the number of $e^\pm$ pairs is reduced in degenerate conditions,
dark photon production by pair annihilation is suppressed at high densities in the core (see \citealt{Rrapaj.Sieverding.ea:2019} for details).
This suppression does not occur in the region at lower densities outside the core, but the temperature of this region
can only facilitate significant emission of the lower-mass dark photons.
Therefore, shell burning phases are accelerated and additional burning episodes with stabilizing effects as described above tend to occur
for $m_A<2\,\mathrm{MeV}$. For heavier dark photons, their production requires energetic $e^\pm$ pairs that can only be provided
by the final core burning stages. Therefore, the assumption that the effect of dark photon cooling is limited to the core burning regions is justified
for $m_A>2\,\mathrm{MeV}$, but the evolution of the outer regions is noticeably affected for smaller dark photon masses.

The stabilizing effect of additional shell burning phases due to the accelerated evolution outside the core cannot be easily captured by a semi-analytical prescription like the one presented in \S\ref{sec:simple_reduction}. Detailed stellar evolution calculations are essential to the discovery and understanding of this effect. 

\subsection{Runaway O Burning}
\label{sec:O-deflagration}
\begin{figure*}
    \centering
    \includegraphics[width=\linewidth]{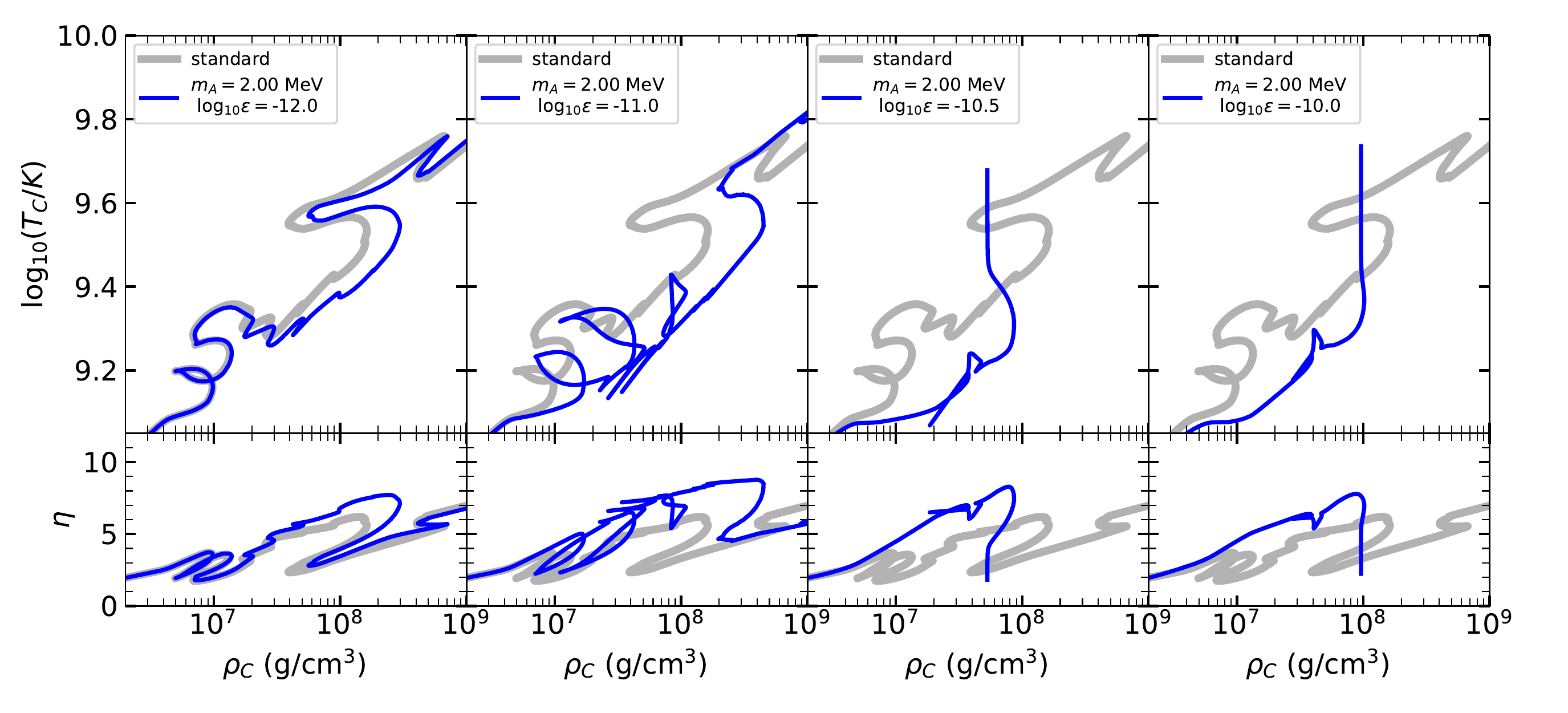}
    \caption{Tracks of central temperature (upper panels) and
    degeneracy parameter $\eta$ (bottom panels) vs. central density
    for models with $m_A=2\,\mathrm{MeV}$ and different values of $\varepsilon$.
    For reference, the fiducial model is shown in gray in all the panels.}
    \label{fig:tracks_2MeV}
\end{figure*}
The stellar models with those dark photon parameters indicated by the red stars in Figure \ref{fig:overview} exhibit a thermonuclear runaway during O~burning. This result may be understood as follows. With the dark photon cooling, the accelerated burning timescales get shorter than the convective and eventually also the hydrodynamic timescales. In addition, the increased electron degeneracy further prevents the star from adjusting its structure
in response to the nuclear energy release. Consequently, the central temperature rises rapidly to $\sim 5$~GK and O is exhausted in a convective region covering less than the innermost $0.1 M_\odot$ of material. Afterwards, O burning propagates outward in a thin shell as a deflagration front.
Figure \ref{fig:tracks_2MeV} shows the evolutionary tracks of $T_\mathrm{C}$ and $\rho_\mathrm{C}$ for $m_A=2\,\mathrm{MeV}$ and 
$\varepsilon=10^{-12}$, $10^{-11}$, $10^{-10.5}$, and $10^{-10}$, respectively, in comparison with the fiducial model. 
The degeneracy parameter $\eta$ is also shown in the bottom panels of this figure.
The onset of the runaway can be seen clearly for the largest two values of $\varepsilon$: $T_\mathrm{C}$ almost vertically increases to $\sim 5\,\mathrm{GK}$ when we stopped the calculations due to the required very small time steps.

Compared to the fiducial model, the tracks for $\varepsilon=10^{-12}$ in Figure \ref{fig:tracks_2MeV} do not show
much deviation, but the density tends to be higher for the same temperature, thereby increasing the degeneracy.
While the track of $T_\mathrm{C}$ vs. $\rho_\mathrm{C}$ for the fiducial model already exhibits a loop, which arises from burning under degenerate conditions 
(see also Figure \ref{fig:standard_track_s15}), the track for $\varepsilon=10^{-11}$ shows two such loops and several backward kinks 
resulting from shell flashes, signifying ignition of shell burning under degenerate conditions. 
For $\varepsilon=10^{-10.5}$, O~burning becomes unstable and the central temperature explosively increases up to $\sim 5$ GK. 
It can be seen from Figure \ref{fig:tracks_2MeV} that in this case the core expands somewhat initially, but the decrease of density fails to stop the runaway heating. 
As $T_\mathrm{C}$ increases rapidly, convection also quickly becomes inefficient to distribute the energy released from the nuclear reactions.
The detailed evolution in this case is discussed in Appendix~\ref{appendix}. We find the same type of behavior as in this case
for even larger values of $\varepsilon$. For example, for $\varepsilon=10^{-10}$, the density response to the rise in temperature 
due to O~burning is barely visible in Figure \ref{fig:tracks_2MeV}, indicating a further acceleration of the runaway process.

While the runaway burning described above results from the interplay of several factors, including the competition of burning timescales with convective and hydrodynamic timescales, as well as partial electron degeneracy, below we present a simple argument for the association of a thermonuclear runaway 
with large values of $\varepsilon$ and provide a conservative estimate for the dark photon luminosity (per unit mass) at which hydrostatic burning is no longer possible. 

\begin{figure}
    \centering
    \includegraphics[width=\linewidth]{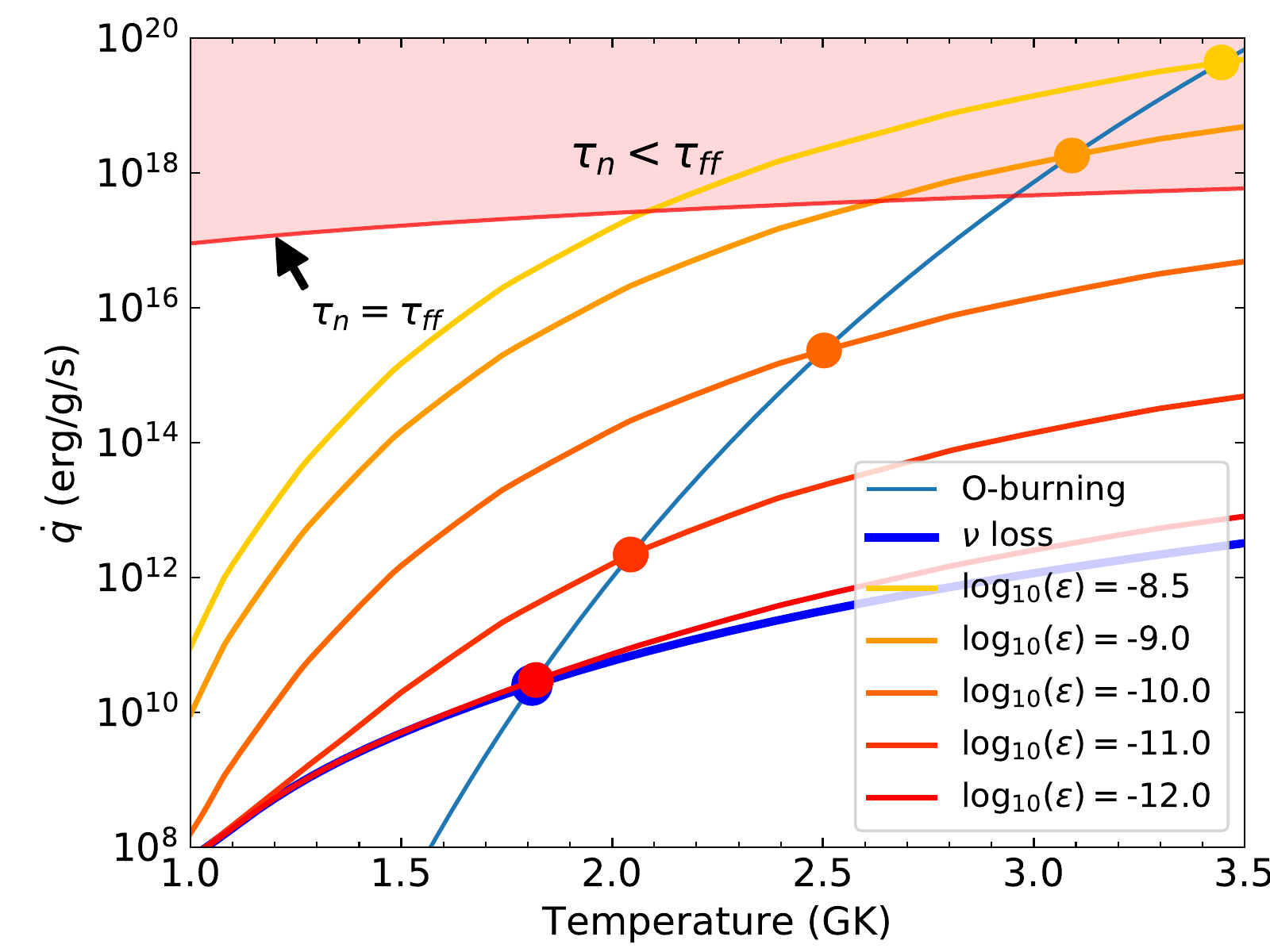}
    \caption{Comparison of the specific energy generation rate for O burning with the loss rates for neutrinos 
    and for dark photons with $m_A=2\,\mathrm{MeV}$. Equating the energy generation and loss rates gives
    the nominal nuclear burning temperatures. For the red shaded area at the top, the nuclear burning timescale $\tau_{\rm n}$
    is shorter than the free-fall timescale $\tau_{\mathrm{ff}}$, and a thermonuclear runaway is expected. See text for detail.}
    \label{fig:e_rates}
\end{figure}

In configurations of stable nuclear burning, the energy release from nuclear reactions is compensated by energy losses, leading to a stable consumption of the nuclear fuel. 
For advanced burning phases without dark photons, the temperature for hydrostatic burning can be estimated by equating the neutrino energy loss and nuclear energy release \citep{Heger.Woosley.ea:2002}. Assuming 
\begin{equation}
\label{eq:rho_of_T}
    \rho=10^6\left(T/\rm{GK}\right)^3\ \rm{g/cm}^3 
\end{equation}
for O burning, the intersection of the O-burning energy generation and the neutrino loss rate gives the typical O-burning temperature of $\approx1.8$~GK
(see blue curves in Figure \ref{fig:e_rates}). 

The additional energy loss due to dark photons changes the temperature for hydrostatic O burning.
Due to the very steep temperature dependence of the nuclear reaction rate, a small change of temperature is sufficient to compensate for a large increase of the energy loss. As shown in Figure \ref{fig:e_rates}, a factor of $\approx 100$ increase of the energy loss for $m_A=2\,\mathrm{MeV}$ and $\varepsilon=10^{-11}$ only slightly increases the nominal O-burning temperature to $\approx 2\,\mathrm{GK}$. Nevertheless, this temperature increase is 
responsible for the speed-up of the evolution discussed in \S\ref{sec:simple_reduction}.
Of course, the burning temperature must increase substantially for strong dark photon couplings.
For example, it reaches $\approx3\,\mathrm{GK}$ for $\varepsilon=10^{-9}$ (see Figure \ref{fig:e_rates}).

Complications arise if the timescale $\tau_\mathrm{n}$ of nuclear burning becomes shorter than the hydrodynamic timescale on which stellar structure can adjust to the temperature change. We take the latter timescale to be the free-fall timescale 
\begin{equation}
    \tau_{\mathrm{ff}}=\frac{1}{\sqrt{G \rho}},
    \label{eq:taudyn}
\end{equation} 
where the density $\rho$ can be estimated from the temperature assuming Eq. (\ref{eq:rho_of_T}).
The timescale of nuclear burning can be estimated as 
\begin{equation}
    \tau_\mathrm{n}=X Q/\dot{q}_{\mathrm{nuc}},
\end{equation} 
where $Q=5\times 10^{17}$~erg/g \citep{Heger.Woosley.ea:2002} is the effective energy release,
$X=0.7$ is the mass fraction of the $^{16}$O fuel, and $\dot{q}_{\mathrm{nuc}}$ is the specific energy generation rate per unit mass. For O burning,
$\tau_{\rm{n}}=\tau_{\rm{ff}}$ occurs at a critical specific energy generation rate of 
\begin{equation}
\label{eq:qcrit}
\dot{q}_{\mathrm{nuc,crit}}\approx 9 \times 10^{16} (T/{\rm GK})^{3/2}\ \rm{erg/g/s}.
\end{equation}
The contour for $\tau_{\rm{n}}=\tau_{\rm{ff}}$ is shown in Figure \ref{fig:e_rates}.
It intersects the curve for the energy release due to O burning at $T\approx 3$~GK
and $\dot{q}_{\rm{nuc,crit}} \approx 4.6\times 10^{17}\,\rm{erg/g/s}$.
At this temperature and the corresponding density from Eq. (\ref{eq:rho_of_T}),
if the specific dark photon energy loss rate exceeds $\dot{q}_{\rm{nuc,crit}}$, i.e.,
\begin{equation}
\label{eq:condition}
 \dot{q}_A(T=3\, \rm{GK}) > 4.6\times 10^{17}\ \rm{erg/g/s},
\end{equation} 
a thermonuclear runaway is expected. As shown in Figure \ref{fig:e_rates},
this result occurs for $m_A=2\,\mathrm{MeV}$ and $\varepsilon=10^{-9}$.

The red shaded region in Figure \ref{fig:overview} indicates
the general parameter space of $m_A$ and $\varepsilon$ for which the condition 
in Eq. (\ref{eq:condition}) is fulfilled.
While the boundary of the region roughly reflects the appearance of the red stars indicating runaway O burning,
the stellar models already exhibit such burning for much weaker couplings for $m_A\lesssim 2$~MeV.
This result may be explained by the complications due to electron degeneracy, which tends to prevent the star from 
reacting to nuclear energy release, and due to the competition of nuclear burning with convection.
Both these complications are ignored in the simple picture assumed for deriving Eq. (\ref{eq:condition}).
At O~ignition, our fiducial model already exhibits some degree of degeneracy with $\eta \sim 2$ (see Figure \ref{fig:tracks_2MeV}). 
When dark photon cooling is included, the temperature of stable burning increases, which requires the stellar core to contract 
to a higher density. Because degeneracy rises more steeply with density than the temperature, values up to $\eta\sim 8$ is reached
in the core at O-burning temperatures when O burning becomes unstable. As the adjustment of the core is slowed down, the O burning 
luminosity exceeds the maximum luminosity that can be transported by convection (see Appendix~\ref{appendix}),
which also favors a local runaway. 

The boundary of the red shaded region in Figure \ref{fig:overview} reflects the dependence of the specific 
dark photon energy loss rate on $m_A$ and $\varepsilon$. It also tends to predict too small threshold values of $\varepsilon$
for runaway burning for higher values of $m_A$. Specifically, for $(m_A/{\rm MeV},\varepsilon)=(4,10^{-7})$,
$(4.5,10^{-6.5})$, $(4.5,10^{-6})$, and $(5,10^{-6})$ that are above the boundary, although O~burning leads to a rapid rise in temperature, 
the dark photon energy loss increases fast enough to avoid a thermonuclear runaway.

\section{Discussion and conclusions}
\label{sec:constraints}
We have studied the impact of extra cooling due to beyond SM particles on the evolution of a $15\,M_\odot$ star.
Specifically, we have implemented the dark photon emission from $e^\pm$ pair annihilation in the stellar evolution code
KEPLER, assuming that these particles decay into other dark sector components, thereby representing an additional 
mechanism of energy loss for the star.
We have considered dark photon masses of $m_A=2\,m_\mathrm{e}$ to 10 MeV and couplings of $\varepsilon=10^{-13}$ to $10^{-6}$,
and found that the dark photon can affect the O~and Si~burning phases. There are three types of potentially observable
effects, which are summarized in Figure \ref{fig:overview}.
\begin{figure}
    \centering
    \includegraphics[width=\linewidth]{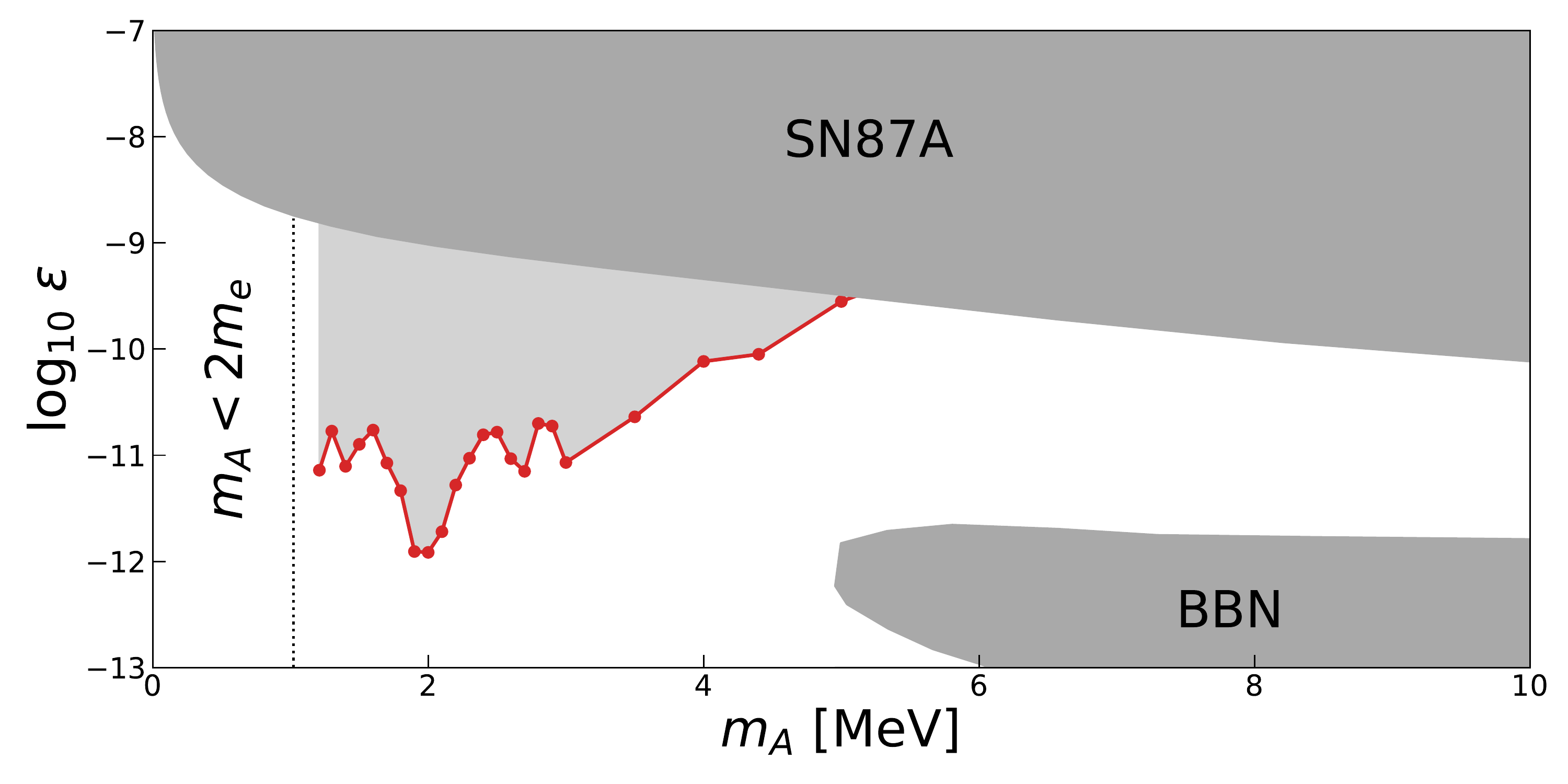}
    \caption{Constraints on dark photon parameters. The light gray region would be excluded if more than
    17 pre-SN neutrinos were detected from a $15\ M_{\odot}$ star at a distance of $500\,\mathrm{pc}$ 
    (the case of $m_A<2\,m_\mathrm{e}$ is not considered).
    The dark gray regions have been excluded by big bang
    nucleosynthesis (BBN, \citealt{Fradette:2014}) and the detection of the neutrinos from SN 1987A
    \citep{Sung.Tu.ea:2019}.}
    \label{fig:exclusion_s15}
\end{figure}
For broad ranges of $m_A$ and $\varepsilon$, the extra cooling allows the star to contract 
faster and increases the temperature at which nuclear burning proceeds in equilibrium with energy losses. 
These effects speed up the burning processes and reduce the number of neutrinos emitted during the last day before core collapse.
We have developed a semi-analytical approach that describes the reduction of the pre-SN neutrino emission for the relevant
ranges of $m_A$ and $\varepsilon$ (see \S\ref{sec:simple_reduction}).

We further find that, due to the density dependence of the emission process, dark photons with $m_A<2\,\mathrm{MeV}$ 
are also produced outside the core, which causes additional shell burning episodes, thereby increasing the pre-SN neutrino 
emission for a very narrow range of parameters (see \S\ref{sec:shell_effects}).

Because extra cooling increases the nominal nuclear burning temperatures, sufficiently strong dark photon couplings are expected 
to produce a thermonuclear runaway (see \S\ref{sec:O-deflagration}). We have found many cases of runaway O burning in our models and 
speculate that thermonuclear SNe may be the outcome in these cases (see Appendix \ref{appendix}). Finding the actual outcome 
and possible observables in these cases, however, requires more refined simulations that can properly follow the propagation 
of narrow nuclear burning fronts. Were dark photon cooling to cause complete disruption of a star, the relevant parameter space 
may be constrained by the observed inventory of neutron stars and stellar black holes, which requires most massive stars to
undergo core-collapse SNe.

The first type of effect, i.e., the suppression of the pre-SN neutrino emission by dark photon cooling, may be constrained by
the detection of such neutrinos. In the most conservative approach, this suppression means that part of the parameter space 
is only consistent with a non-detection of pre-SN neutrinos. Therefore, any positive detection of a pre-SN neutrino signal 
from a well-understood progenitor excludes this part of the dark photon parameter space.
For illustration, we take our $15\, M_{\odot}$ model to explode as a core-collapse SN at a distance of $500\,\mathrm{pc}$. 
Assuming the normal neutrino mass hierarchy, we can find the values of ($m_A, \varepsilon$) that reduce the pre-SN neutrino 
emission sufficiently to give an expected number of events less than the background count of about $17$ events, in contrast
to about 50 events without dark photon cooling.
The corresponding parameter space is bounded by the red curve in Figure \ref{fig:exclusion_s15}. The detection a number of events significantly above the background level
would exclude this part of the parameter space, if the progenitor is similar to the model we have studied here.
Note that a small fraction of this parameter space may lead to a thermonuclear runaway, and therefore, could be constrained
by the observation of a core-collapse rather than thermonuclear SN.

Pre-SN neutrinos have not been detected yet.
Our results, however, indicate that their observation in the future may offer a unique probe of not only the dark photon 
but also other dark matter particles that are efficiently produced in the temperature regime of $\sim1$--$10\,\mathrm{GK}$
during stellar evolution.

\acknowledgments
This work was supported in part by the US Department of Energy [DE-FG02-87ER40328 (UM)].
Calculations were carried out at the Minnesota Supercomputing Institute.
ER was supported by the NSF (PHY-1630782) and the Heising-Simons Foundation (2017-228).  GG acknowledges support from the Academia Sinica by Grant No. AS-CDA-109-M11.
We thank Alexander Heger for providing access to the KEPLER code.

\appendix
\section{A model with Runaway O Burning}
\label{appendix}
\begin{figure*}
\gridline{\fig{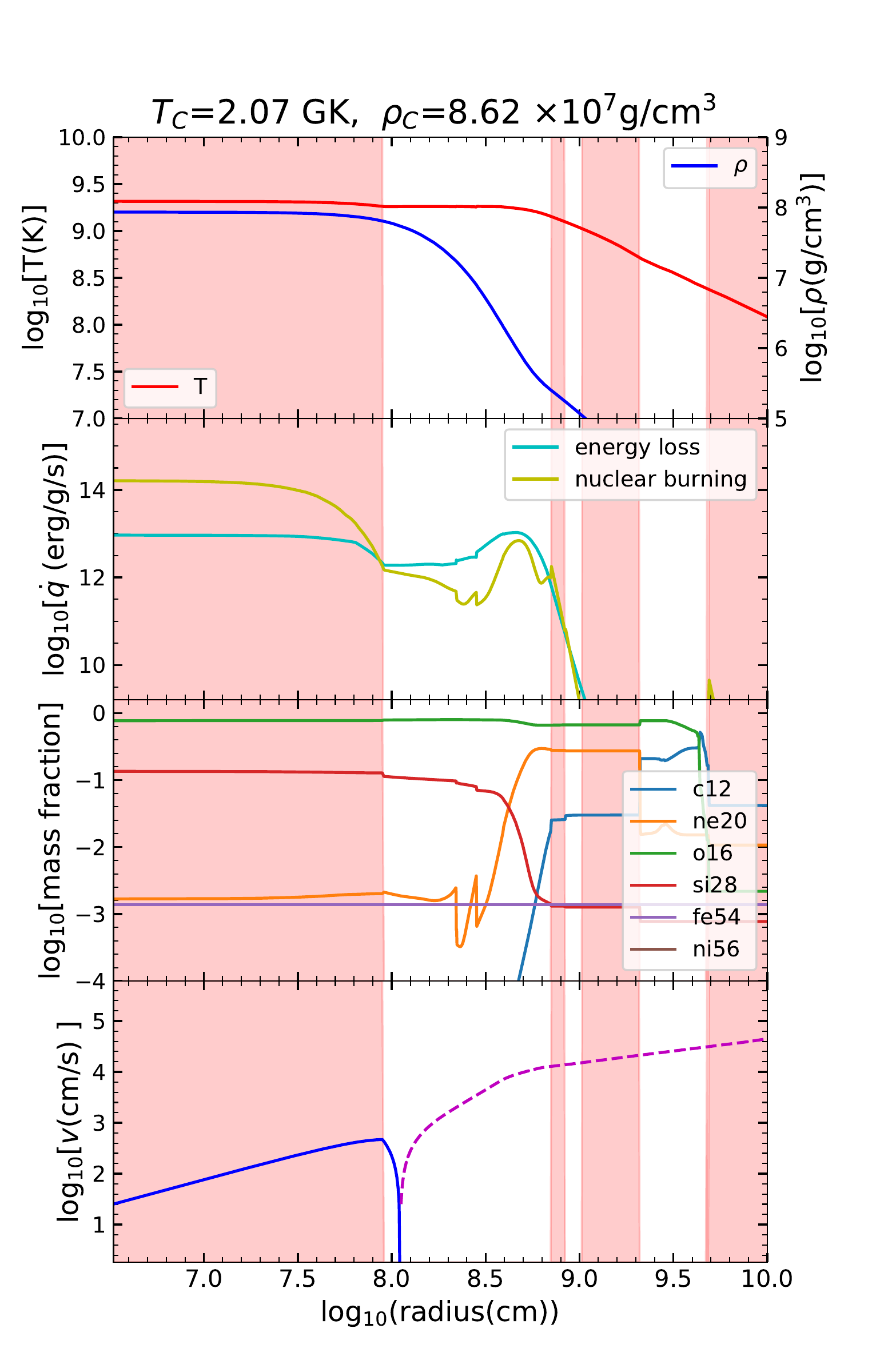}{0.32\textwidth}{(a)}
          \fig{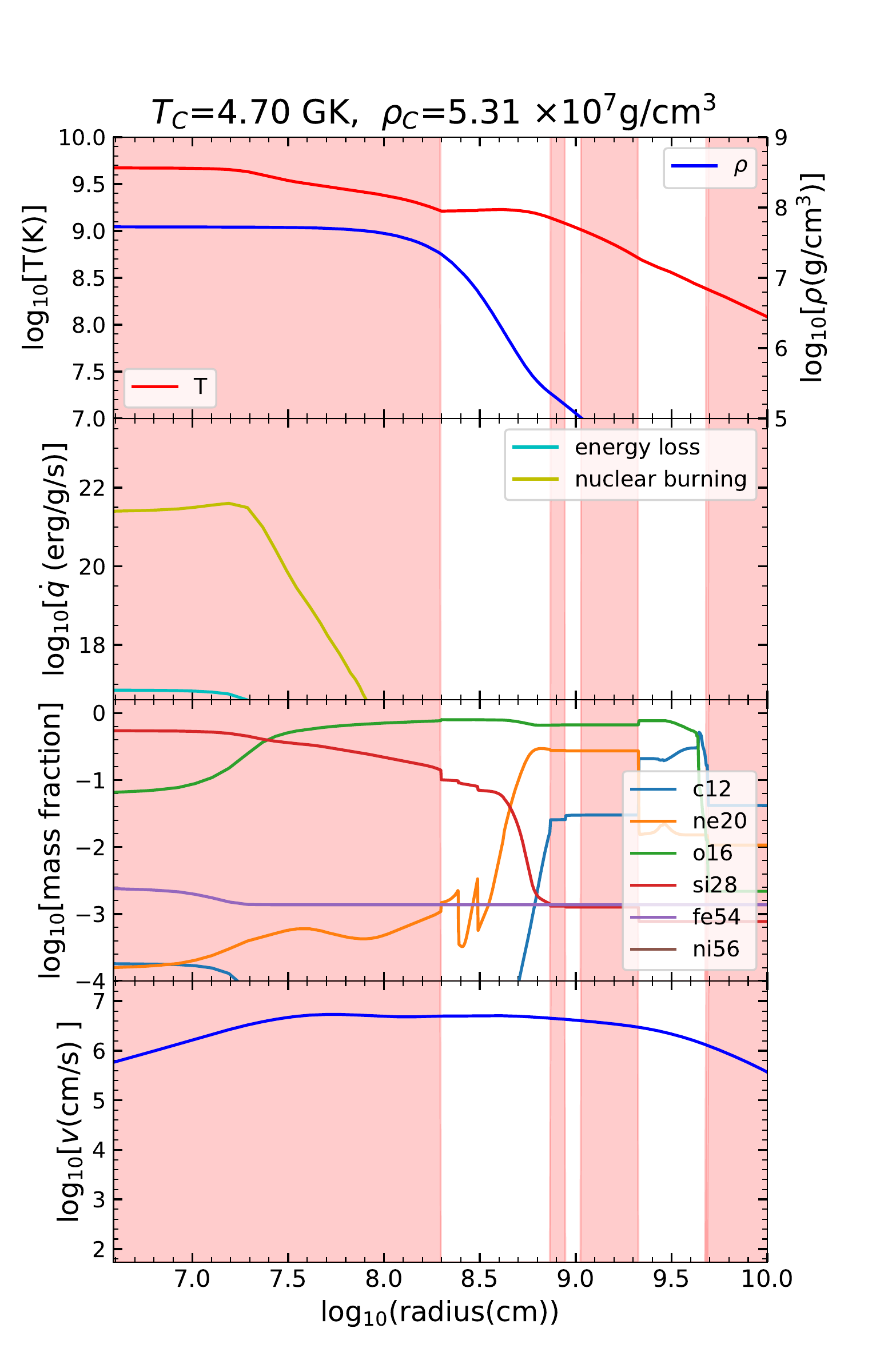}{0.32\textwidth}{(b)}
          \fig{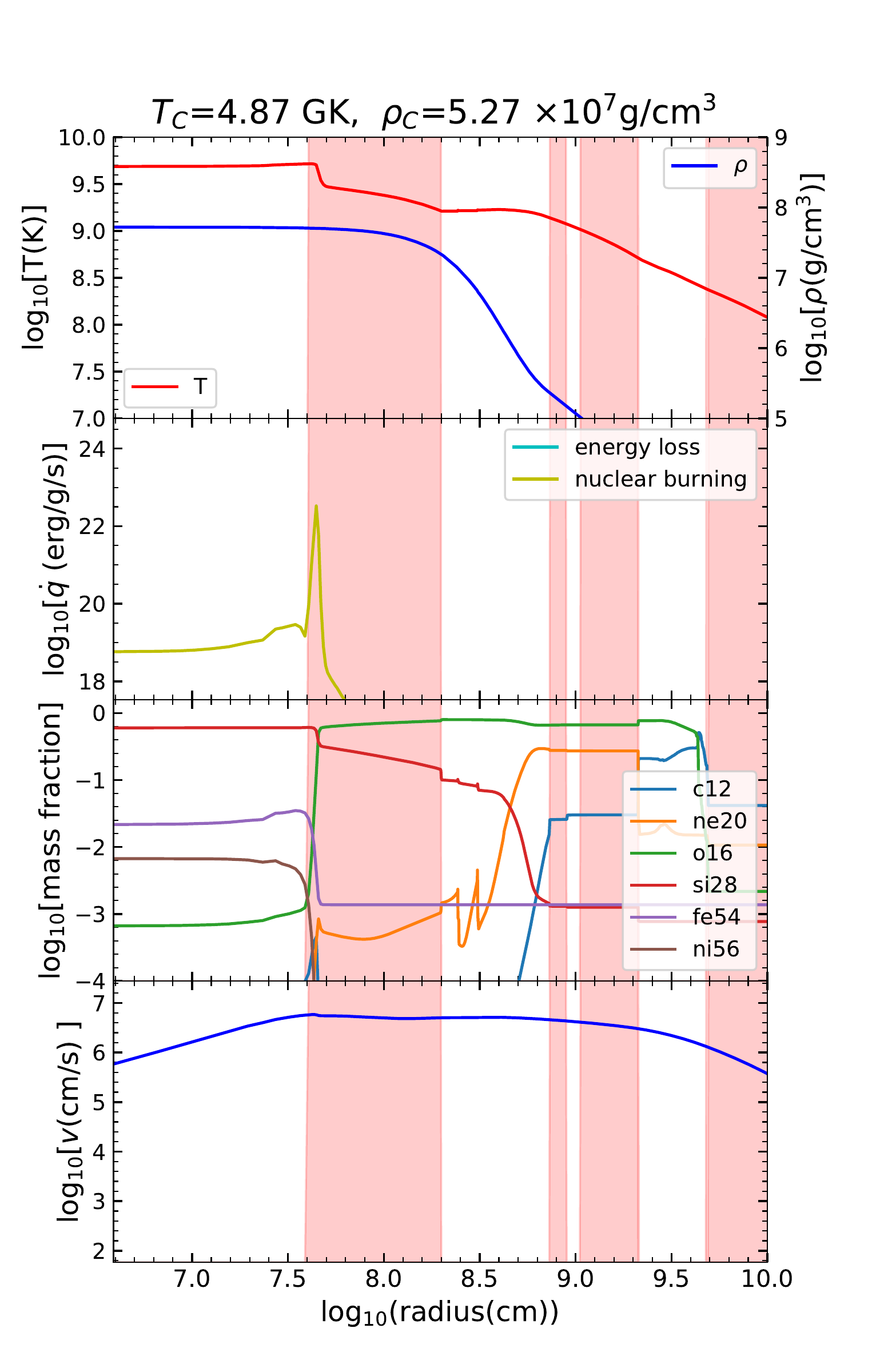}{0.32\textwidth}{(c)}}
          \caption{Snapshot profiles of the stellar core for the model with $m_A=2\,\mathrm{MeV}$ and $\varepsilon=10^{-10.5}$. Time progresses from left to right and convective regions are shown as red shaded areas. The top row shows the temperature and density profiles, the second row shows the nuclear energy release and combined neutrino and dark photon loss, the third row shows the mass fractions of the key isotopes, and the bottom row shows the velocity profile, where positive (outward) velocities are displayed as solid curves and negative (infall) velocities as dashed curves.
          Panel (a) shows the beginning of O~burning when the energy release from nuclear reactions just exceeds the combined neutrino and dark photon loss. Panel (b) shows a snapshot 275~s after panel (a), when the energy generation at the center starts to decrease as O is exhausted. The high luminosity cannot be distributed efficiently by convection and only the region of the inner 300 km finishes O burning. Panel (c) is 82~s after panel (b) and shows that the burning is compressed into a narrow burning front, where the temperature jumps from $3\,\mathrm{GK}$ to $5\,\mathrm{GK}$. }
          \label{fig:profiles}
\end{figure*}

In order to understand the phenomenon of runaway O burning,
we investigate the model for $m_A=2$~MeV and $\varepsilon=10^{-10.5}$ in more detail.
Figure \ref{fig:profiles} shows a sequence of snapshots from the evolution of this model.
The top row shows the temperature and density profiles, the second row from the top shows
the nuclear energy release and combined energy loss due to neutrinos and dark photons, 
the third row shows the mass fractions of the major isotopes to illustrate the progress of nuclear burning,
and the bottom row shows the velocity profile, where positive (outward) velocities are shown as solid 
curves and negative (infall) velocities are represented by dashed curves. 
The red shaded regions are convective zones. Time runs from panel (a) on the left to panel (c) on the right, spanning $357\,\mathrm{s}$. 
Panel (a) shows the onset of O~burning, which is confined to a convective region reaching up to 1000 km and encompassing about 0.1 $M_\odot$ of material. At $T\approx 2\,\mathrm{GK}$ and $\rho\approx 8\times10^7\,\mathrm{g/cm}^3$, the energy release from nuclear reactions exceeds the combined neutrino and dark photon loss. In response to the excess heat, positive velocities develop, indicating the onset of core expansion. Panel (b), however, shows that O~burning proceeds much faster than the adjustment of stellar structure. The fuel in the center is almost depleted and the temperature has risen to $4.7\,\mathrm{GK}$ while the density has only been reduced by about $40\,\%$ compared to panel (a). 

At a temperature above $4\,\mathrm{GK}$, the luminosity due to O burning exceeds the maximum value that can be transported by convection \citep{Woosley.Heger.ea:2015}
\begin{equation}
  L_{\mathrm{max}} \approx  4 \pi r^2 \rho v_{\mathrm{conv}} f C_\mathrm{P} T, 
    \label{eq:lmax}
\end{equation}
where $v_{\mathrm{conv}}$ is the convective velocity, $C_\mathrm{P} T$ is the thermal energy content, and $f\ll 1$ 
indicates the efficiency of convection to remove the internal energy. With $f =0.1$ we find $L_\mathrm{max}<10^{50}\mathrm{erg/s}$, corresponding to an energy generation rate of $\dot{q}_{\mathrm{max}} \approx 5\times 10^{17}\,\mathrm{erg/g/s}$ for the conditions of O~burning assuming an initial core mass of $0.1\,M_\odot$. The nuclear energy release in panel (b) exceeds this luminosity already by four orders of magnitude and convection cannot distribute the heat and the fuel throughout the convective region on the burning timescale. Therefore, O is only depleted in the innermost $300\,\mathrm{km}$ despite that the convective zone reaches out to $2000\,\mathrm{km}$.

Without efficient convection, the very rapid burning gets confined into a narrow burning front at the bottom of the nominally convective layer, as shown in panel (c) of Figure \ref{fig:profiles}. The change of composition and continuing energy generation also show that Si burning immediately follows because of the high temperature of the central region.
In our model, the burning region consists only of a few zones and the fuel in one zone is consumed before the zone above it ignites. The propagation of the burning depends critically on the heat transport and would need to be described by a model for flame propagation. 
For a very narrow burning front, multi-dimensional effects are important for the propagation of such a deflagration \cite{Fryxell.Woosley:1982} and the outcome is highly sensitive to turbulence and mixing at the boundary of the burning front \citep{Jones.Hirschi.ea:2013,Jones.Roepke.ea:2016}.
The bottom row of Figure \ref{fig:profiles} also shows that significant outward velocities develop but remain subsonic. 
The above situation cannot be adequately simulated by our model due to the lack of resolution and the employed small reaction network and mixing-length treatment of convection. Therefore, our results on the final outcome are speculative.

We have followed the calculation of our model further and find that the shell burning steepens into a supersonic shock once it reaches the steep density gradient at the upper edge of the convective zone around the radius of 2000 km, where Ne burning also provides additional energy. The outward velocity reaches several 1000 km/s, which takes the pressure off the core, eventually facilitating a rapid expansion and cooling. The $T_\mathrm{C}$ drops below 0.1 GK and $\rho_\mathrm{C}$ below $10^5$ g/cm$^3$. When the shock reaches the H envelope we find that the kinetic energy of $3.6\times 10^{50}$ erg exceeds the gravitational binding energy of $1.4\times 10^{50}$ erg for the star. The shock may thus disrupt the whole star or at least unbind a significant fraction of the H envelope and lead in either case to an optical transient. Even if the core eventually collapses, the runaway O burning would delay the collapse by many hours up to days. Here we caution again that our model is probably inadequate to describe this situation accurately and a better treatment is needed to predict potentially observable signatures.

\bibliographystyle{aasjournal}

\end{document}